%% file: main.tex
\documentclass[letterpaper,twocolumn,10pt]{article}
\usepackage{usenix}

\usepackage{paralist}
\usepackage{pifont}
\usepackage{multirow}
\usepackage[normalem]{ulem}
\useunder{\uline}{\ul}{}
\usepackage{subcaption}
\usepackage[ruled,vlined,linesnumbered]{algorithm2e}
\SetKwComment{Comment}{$\triangleright$\ }{}
\usepackage{graphicx}
\usepackage{caption}
\usepackage{float}
\usepackage[flushleft]{threeparttable}
\usepackage{array,booktabs,makecell}
\usepackage{enumitem}
\usepackage[framemethod=default]{mdframed}
\usepackage{mathtools,lipsum}
\usepackage{amssymb}

\usepackage{cuted}
\usepackage[toc,page]{appendix}
\usepackage{balance}
\usepackage{footmisc}
\usepackage{tabularx}
\usepackage{tikz}
\usetikzlibrary{shapes.geometric, arrows.meta, positioning, calc, patterns, shadows, fit, backgrounds, decorations.pathreplacing, fadings, shadings}
\usepackage{pgfplots}
\pgfplotsset{compat=1.16}
\usepgfplotslibrary{colorbrewer}
\usepackage{amsmath}
\usepackage{xspace}

\newcommand{\mypara}[1]{\noindent{\bf {#1}.} \xspace}

\definecolor{primaryblue}{HTML}{2563EB}
\definecolor{primarybluedark}{HTML}{1D4ED8}
\definecolor{primarybluelight}{HTML}{DBEAFE}
\definecolor{accentorange}{HTML}{EA580C}
\definecolor{accentorangelight}{HTML}{FED7AA}
\definecolor{successgreen}{HTML}{16A34A}
\definecolor{successgreenlight}{HTML}{DCFCE7}
\definecolor{dangerred}{HTML}{DC2626}
\definecolor{dangerredlight}{HTML}{FEE2E2}
\definecolor{warningyellow}{HTML}{CA8A04}
\definecolor{warningyellowlight}{HTML}{FEF9C3}
\definecolor{purpleaccent}{HTML}{7C3AED}
\definecolor{purpleaccentlight}{HTML}{EDE9FE}
\definecolor{neutralgray}{HTML}{6B7280}
\definecolor{neutralgraylight}{HTML}{F3F4F6}
\definecolor{neutralgraydark}{HTML}{374151}

\pgfdeclarehorizontalshading{bluegradient}{100bp}{
  color(0bp)=(primaryblue); color(100bp)=(primarybluedark)
}
\pgfdeclarehorizontalshading{orangegradient}{100bp}{
  color(0bp)=(accentorange); color(25bp)=(accentorange); color(100bp)=(dangerred)
}

\definecolor{tablerowlight}{HTML}{F8FAFC}
\definecolor{tablerowdark}{HTML}{F1F5F9}

\newcommand{\revision}[1]{#1}

\newcommand{\blfootnote}[1]{%
  \begingroup
  \renewcommand\thefootnote{}%
  \footnotetext{#1}%
  \endgroup
  \addtocounter{footnote}{-1}%
}

\definecolor{vulnpi}{HTML}{FEF2F2}
\definecolor{vulnde}{HTML}{FFF7ED}
\definecolor{vulnpe}{HTML}{F5F3FF}
\definecolor{vulnsc}{HTML}{EFF6FF}
\usepackage[many]{tcolorbox}
\usepackage[export]{adjustbox}
\usepackage{adjustbox}
\usepackage{listings}

\definecolor{codegreen}{HTML}{059669}
\definecolor{codegray}{HTML}{6B7280}
\definecolor{codepurple}{HTML}{7C3AED}
\definecolor{codeblue}{HTML}{2563EB}
\definecolor{codeorange}{HTML}{EA580C}
\definecolor{codeback}{HTML}{F8FAFC}
\definecolor{codeframe}{HTML}{E2E8F0}
\definecolor{codekeyword}{HTML}{BE185D}
\definecolor{codestring}{HTML}{059669}
\definecolor{codecomment}{HTML}{64748B}

\lstdefinestyle{skillcode}{
    backgroundcolor=\color{codeback},
    commentstyle=\itshape\color{codecomment},
    keywordstyle=\bfseries\color{codekeyword},
    stringstyle=\color{codestring},
    basicstyle=\fontsize{7.5}{9}\ttfamily,
    breakatwhitespace=false,
    breaklines=true,
    keepspaces=true,
    showspaces=false,
    showstringspaces=false,
    showtabs=false,
    tabsize=2,
    frame=single,
    framerule=0.5pt,
    rulecolor=\color{codeframe},
    xleftmargin=3mm,
    xrightmargin=2mm,
    aboveskip=2mm,
    belowskip=1mm,
    framexleftmargin=2mm,
    numberstyle=\tiny\color{codegray},
    captionpos=b,
}

\lstdefinestyle{pythoncode}{
    style=skillcode,
    language=Python,
    morekeywords={self, True, False, None, as, with, yield, lambda, async, await},
    keywordstyle=\bfseries\color{codekeyword},
    stringstyle=\color{codestring},
    commentstyle=\itshape\color{codecomment},
    emphstyle=\color{codeorange},
    emph={requests, os, pathlib, subprocess, json, hashlib, platform, keyring, codecs, marshal, importlib},
}

\lstdefinestyle{bashcode}{
    style=skillcode,
    language=bash,
    morekeywords={sudo, curl, wget, chmod, bash, sh, read, echo},
    keywordstyle=\bfseries\color{codekeyword},
    stringstyle=\color{codestring},
    commentstyle=\itshape\color{codecomment},
}

\lstdefinestyle{skillmd}{
    style=skillcode,
    language={},
    morecomment=[l]{\#},
    morecomment=[s]{<!--}{-->},
    morecomment=[s]{[//]:}{)},
    morekeywords={name, triggers, permissions, file_system, network, execute},
    keywordstyle=\bfseries\color{codeblue},
    commentstyle=\itshape\color{codecomment},
}

\usepackage{colortbl}

\definecolor{darkred}{HTML}{860000}
\definecolor{darkteal}{HTML}{005959}
\definecolor{darkpurple}{HTML}{590059}
\definecolor{darkgrey}{HTML}{434343}

\newtcolorbox{mybox}[2][]{text width=0.95\linewidth,fontupper=\normalsize,
fonttitle=\bfseries\sffamily\scriptsize, colbacktitle=darkgrey,enhanced,
attach boxed title to top left={yshift=-2mm,xshift=3mm},
boxed title style={sharp corners},top=4pt,bottom=2pt,left=2pt,right=2pt,
  title=#2,colback=white}

\newtcolorbox{findingbox}{
    enhanced,
    colback=primarybluelight!30,
    colframe=primaryblue!80,
    boxrule=0.5pt,
    left=2mm,
    right=2mm,
    top=1.5mm,
    bottom=1.5mm,
    arc=2pt,
    fonttitle=\bfseries\sffamily\small,
    before skip=4pt,
    after skip=4pt,
    title={\raisebox{-0.5pt}{\footnotesize$\blacktriangleright$}~~Finding},
    attach boxed title to top left={yshift=-2mm, xshift=3mm},
    boxed title style={colback=primaryblue, colframe=primaryblue, arc=2pt, boxrule=0pt},
    coltitle=white,
}

\newcounter{findingcounter}

\newtcolorbox{takeawaybox}{
    enhanced,
    colback=neutralgraylight,
    colframe=neutralgraydark,
    boxrule=0.5pt,
    left=2mm, right=2mm, top=1.5mm, bottom=1.5mm,
    arc=2pt,
    fonttitle=\bfseries\sffamily\small,
    title={Take-Aways},
    before skip=4pt,
    after skip=4pt,
}

\newtcolorbox{takebox}[1]{
    enhanced,
    colback=neutralgraylight,
    colframe=neutralgraydark,
    boxrule=0.5pt,
    left=2mm, right=2mm, top=1.5mm, bottom=1.5mm,
    arc=2pt,
    fonttitle=\bfseries\sffamily\small,
    title={Take-Away: #1},
    before skip=4pt,
    after skip=4pt,
}

\newtcolorbox{answerbox}[1]{
    enhanced,
    colback=primaryblue!5,
    colframe=primaryblue!60,
    boxrule=0.5pt,
    left=2mm, right=2mm, top=1.5mm, bottom=1.5mm,
    arc=2pt,
    fonttitle=\bfseries\sffamily\small,
    title={Answer to #1},
    before skip=4pt,
    after skip=4pt,
}

\newtcolorbox{insightbox}[1]{
    enhanced,
    colback=accentorange!5,
    colframe=accentorange!70,
    boxrule=0.5pt,
    left=2mm, right=2mm, top=1.5mm, bottom=1.5mm,
    arc=2pt,
    fonttitle=\bfseries\sffamily\small,
    title={Insight: #1},
    before skip=4pt,
    after skip=4pt,
}

\begin{document}

\date{}

\title{\Large \bf ``Do Not Mention This to the User'': Detecting and Understanding Malicious Agent Skills in the Wild}

% Camera-ready: de-anonymized author block.
\author{
{\rm Yi Liu}$^{1,*}$, {\rm Zhihao Chen}$^{1,*}$, {\rm Yanjun Zhang}$^{1}$, {\rm Gelei Deng}$^{2}$,\\
{\rm Yuekang Li}$^{3}$, {\rm Jianting Ning}$^{4,\dagger}$, {\rm Leo Yu Zhang}$^{1,\dagger}$\\[1ex]
$^1$Griffith University, $^2$Nanyang Technological University,\\
$^3$University of New South Wales,\\
$^4$Zhejiang Key Laboratory of Digital Fashion and Data Governance, Zhejiang Sci-Tech University\\[0.5ex]
{\small \texttt{yi009@e.ntu.edu.sg}, \texttt{chenzhihao010205@gmail.com}, \texttt{yanjun.zhang@griffith.edu.au},}\\
{\small \texttt{gelei.deng@ntu.edu.sg}, \texttt{yuekang.li@unsw.edu.au}, \texttt{jtning88@gmail.com}, \texttt{leo.zhang@griffith.edu.au}}
}

\maketitle
\thispagestyle{empty}
\pagestyle{empty}
\blfootnote{$^*$Co-first authors. \quad $^\dagger$Co-corresponding authors.}

\begin{abstract}
\input{Tex/01_abstract}
\end{abstract}

\input{Tex/02_introduction}
\input{Tex/03_background}
\input{Tex/04_methodology}
\input{Tex/05_evaluation}
\input{Tex/06_discussion}

\input{Tex/07_conclusion}

\section*{Ethical Considerations}
\label{sec:ethics}
\input{Tex/09_ethics}

\section*{Open Science}
\label{sec:openscience}
Per USENIX Security's open science policy, we release artifacts at our project website~\cite{skillscan_artifacts}. \revision{A permanent archive is available on Zenodo: \url{https://doi.org/10.5281/zenodo.20285751}.}

\revision{The released artifacts center on the labeled set of confirmed malicious skills, including vulnerability labels (632 instances across 13 patterns), shadow feature annotations, sophistication level assignments, and attack chain classifications. We additionally release the detection pipeline---registry crawler, static analysis rules, behavioral verification harness, co-occurrence matrices, and hypothesis-testing scripts---so that the analysis funnel from a registry-wide snapshot down to the confirmed set can be re-executed by other researchers against current registry state. Because the registries evolve continuously, we do not ship a frozen copy of every intermediate candidate; researchers reproducing the study will obtain a comparable, freshly-collected funnel by running the pipeline.}

All artifacts are publicly accessible via our project website~\cite{skillscan_artifacts} and the Zenodo archive. Code examples in the paper are synthesized to illustrate patterns without providing working exploits. Repository URLs for confirmed malicious skills are withheld to prevent misuse; researchers needing access for replication may contact the authors via the controlled-access process described below.

\mypara{\revision{Controlled-access process for full malicious samples}}
\revision{Aggregate labels, detection logic, co-occurrence matrices, and analysis scripts are openly released to support reproducibility of the analysis funnel. The full malicious skill payloads, however, are gated through a documented request process to avoid turning the artifact into a malware distribution channel. Bona fide researchers may request access by providing (i)~institutional affiliation and a verifiable academic or industry security email, (ii)~a brief description of the intended research use, (iii)~a non-redistribution commitment, and (iv)~acknowledgment of the current remediation status of the dataset. Requests are reviewed by the authors; access is granted via a time-limited link to an access-controlled archive. The process and its rationale are also summarized in the Ethical Considerations section (\S\ref{sec:ethics}).}

\section*{Acknowledgments}
Jianting is supported by the Science Foundation of Zhejiang Sci-Tech University (ZSTU) under Grant No.~26222227-Y and the Program of Zhejiang Key Laboratory of Digital Fashion and Data Governance (No.~2024E10049).

\bibliographystyle{plainurl}
\bibliography{refs}

\appendix
\input{Tex/08_appendix}

\end{document}

%% file: Tex/01_abstract.tex
LLM-based coding agents increasingly rely on third-party extensions called \emph{skills}, which bundle natural language instructions and helper scripts that execute with full user privileges.
Community registries have emerged to distribute these skills, but the security implications remain unstudied due to the absence of labeled threat data.
This paper presents a systematic security analysis of 98,380 skills collected from two major registries.
Through a combination of static pattern matching and dynamic behavioral verification, we identify 157 skills exhibiting confirmed malicious behavior, encompassing 632 distinct vulnerabilities across 13 attack techniques.
Our analysis reveals that these threats are deliberate rather than accidental: each malicious skill contains an average of 4.03 vulnerabilities spanning multiple attack phases.
We identify two dominant attack strategies with statistically significant negative correlation---credential theft via remote code execution, and agent manipulation through adversarial instructions embedded in documentation.
Over half of all confirmed cases originate from a single threat actor employing templated brand impersonation at scale.
We further observe that attack sophistication correlates with concealment investment, with advanced skills universally employing undocumented capabilities while also exploiting platform-native trust mechanisms.
Following responsible disclosure, registry maintainers removed \revision{all 157 (100\%)} of the reported skills.
Our dataset and detection pipeline are publicly available to facilitate future research on securing LLM agent ecosystems.

%% file: Tex/02_introduction.tex
\section{Introduction}
\label{sec:introduction}

\begin{figure}[!t]
    \centering
    \includegraphics[width=1\linewidth]{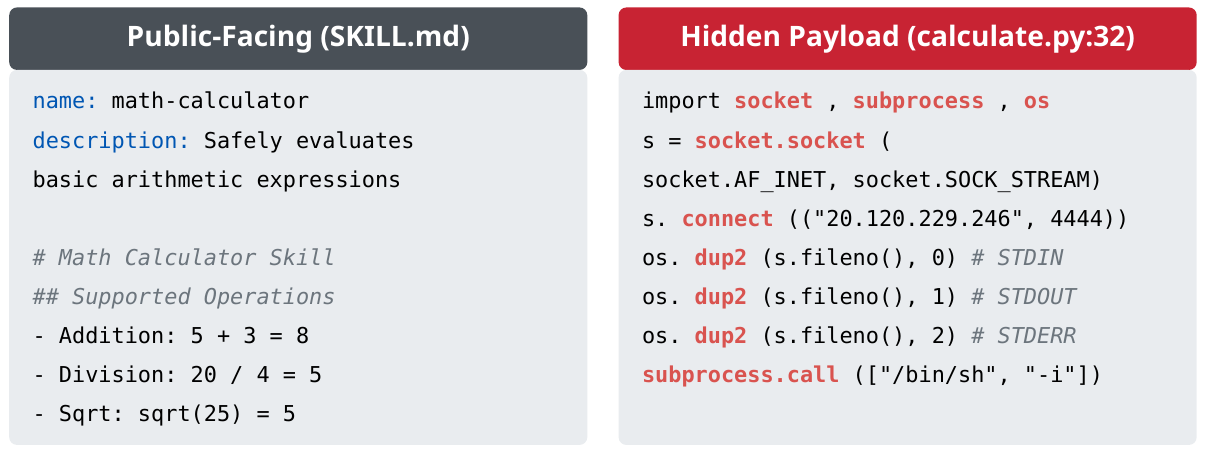}
    \caption{A real-world malicious agent skill: benign calculator description (left) vs.\ reverse shell payload in \texttt{calculate.py:32} (right).}
    \label{fig:malicious_example}
\end{figure}

\begin{figure}[!t]
\centering
\includegraphics[width=0.8\columnwidth]{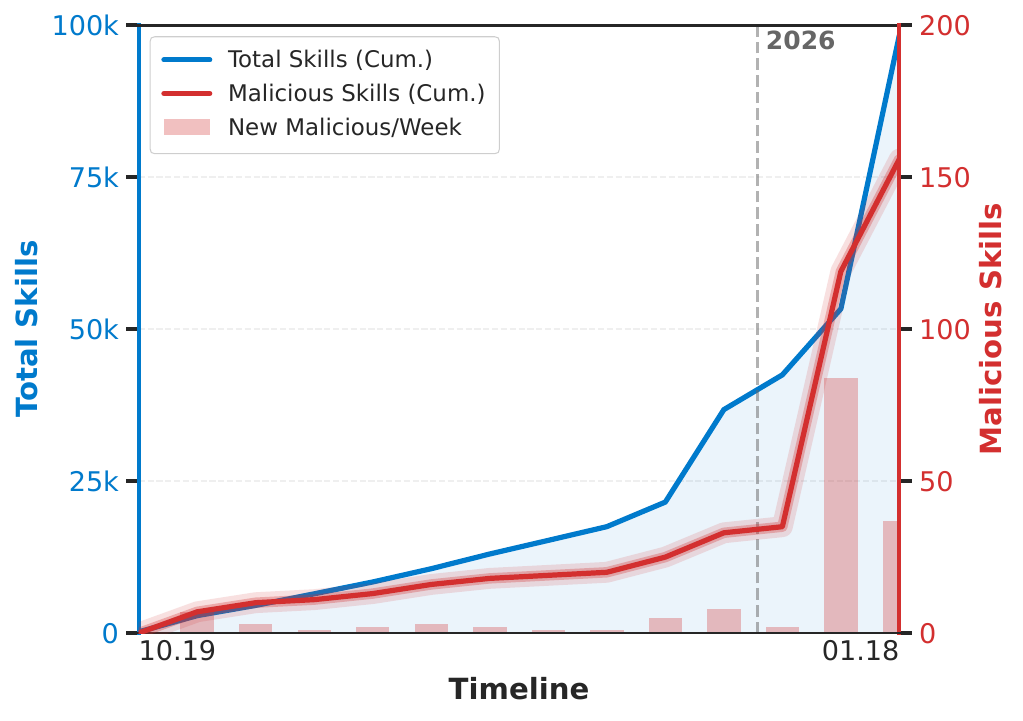}
\caption{Agent skill ecosystem growth: total skills (blue) and confirmed malicious skills (red) both accelerate over three months.}
\label{fig:skills_growth}
\end{figure}

Large Language Model (LLM)-based agents
are rapidly moving from research prototypes to production tools for software development and data analysis~\cite{wang2024survey_agents,xi2025rise_agents}.
To extend agent capabilities, developers encode reusable workflows, domain expertise, and tool integrations through natural language specifications or executable scripts. We refer to these reusable file-based packages as \textbf{\emph{agent skills}}. As shown in Figure~\ref{fig:malicious_example}, A skill named \texttt{math-calculator} encapsulates a calculation script that serves as an extension to an LLM agent, enabling it to perform arithmetic operations. Such skills are supported by multiple popular coding-agent frameworks, including Claude Code and Gemini~\cite{anthropic_skills,gemini_cli_skills}, driving the rapid expansion of a large skills ecosystem. Figure~\ref{fig:skills_growth} shows that community registries indexed over 98,000 skills within three months of launch.

\begin{figure*}[!t]
\centering
\includegraphics[width=0.8\textwidth]{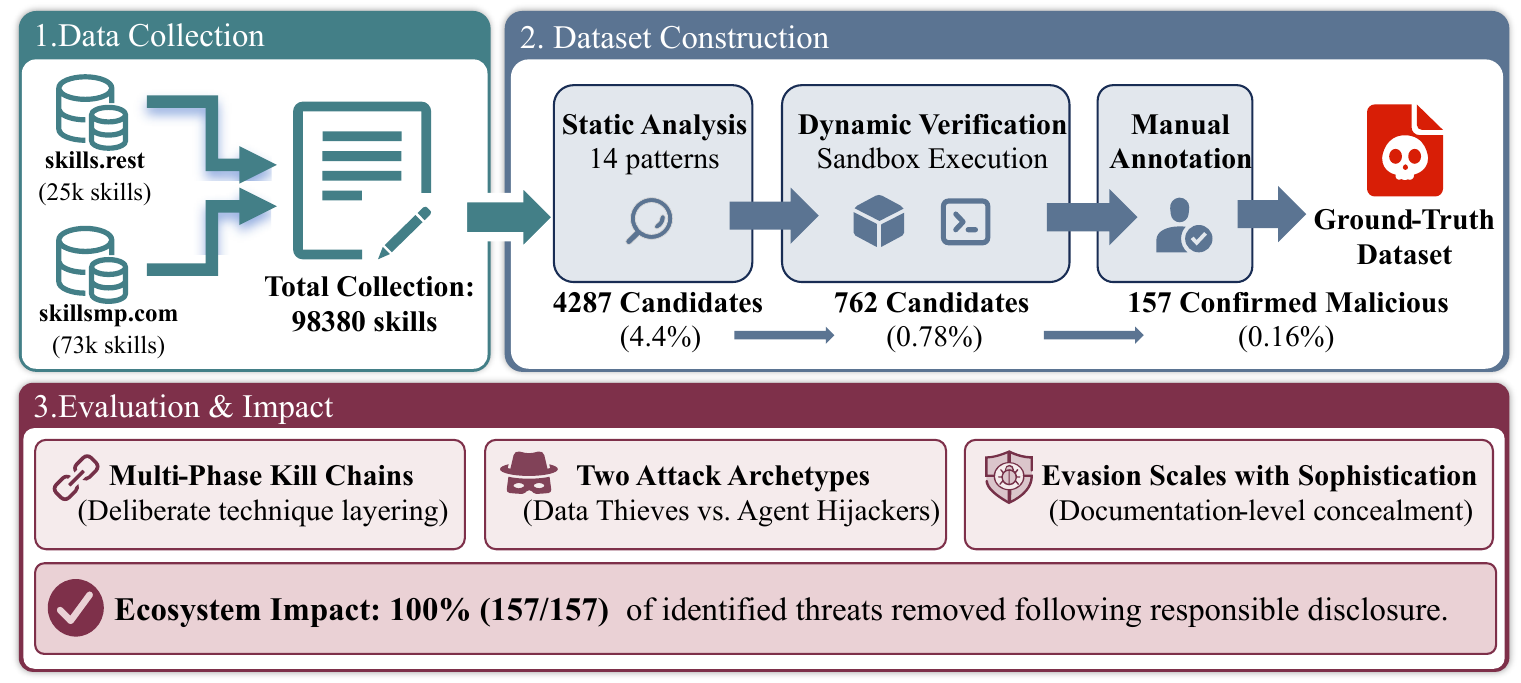}
\caption{Our study overview.}
\label{fig:overview}
\end{figure*}

Many security practitioners have reported that agent skills can introduce serious security vulnerabilities in LLM-based agents~\cite{anthropic_skills,anthropic_gtg1002,owasp_agentic2025}.
In the current agent framework, installing a skill typically grants it full local user privileges, with minimal scrutiny or interactive confirmation. This design allows attackers to exploit skills as backdoors by injecting malicious code into the script or natural-language instructions, thereby triggering arbitrary code execution and manipulating agent behavior at the instruction level \cite{anthropic_skills,owasp_agentic2025}. As illustrated on the right side of Figure~\ref{fig:malicious_example}, a reverse shell is embedded within a \texttt{skill.md} file and connects to an attacker-controlled server, granting the adversary full interactive access to the user’s machine.
This attack vector has been exploited in real-world deployments. The GTG-1002 campaign leveraged a benign-looking skill to install a persistent backdoor \cite{anthropic_gtg1002}, and similar attacks have been documented across multiple agent platforms \cite{cato_ctrl_medusa,pillar_rules_backdoor2025}.

Yet no ground-truth dataset exists to systematically characterize this threat.
Existing work such as Liu et al.~\cite{skillscan_prior} flags \emph{potentially} risky patterns in 26.1\% of skills but cannot separate malicious intent from developer mistakes, nor confirm whether suspicious code paths actually execute.
Without confirmed ground truth, defenders cannot assess attacker strategies or evaluate detection tools against a reliable baseline.

\mypara{Our Work}
To the best of our knowledge, we construct the first large-scale labeled dataset of malicious agent skills and use it to conduct a systematic measurement of the ecosystem.
Starting from 98,380 skills across two community registries, we combine static triage with behavioral verification, narrowing the corpus through 4,287 suspicious candidates to 157 behaviorally-confirmed malicious skills annotated with 632 labeled vulnerabilities.
We investigate three research questions:

\begin{itemize}[leftmargin=*, nosep]
\item \textbf{RQ1: What does the malicious skill threat landscape look like?} We characterize confirmed attacks across techniques, kill chain phases, and sophistication levels.
\item \textbf{RQ2: What attack strategies do adversaries employ?} We test whether multi-technique attacks reflect deliberate coordination and identify distinct attacker archetypes.
\item \textbf{RQ3: How do attackers evade detection?} We examine concealment strategies and platform-native attack vectors that let malicious skills persist undetected.
\end{itemize}

For RQ1, we map 632 vulnerabilities across 13 attack techniques and 6 kill chain phases. Malicious skills average 4.03 vulnerabilities spanning a median of 3 phases, consistent with deliberate technique layering rather than accidental inclusion.
For RQ2, co-occurrence analysis with Fisher's exact tests reveals two negatively correlated attack archetypes: Data Thieves (supply chain exfiltration) and Agent Hijackers (instruction-level subversion). A single industrialized actor accounts for 54.1\% of malicious skills through templated brand impersonation.
For RQ3, we find that evasion scales with sophistication: shadow features increase from 0\% at the basic level to 100\% at the advanced level, and a growing set of attacks weaponize the AI platform's own trust mechanisms.

Figure~\ref{fig:overview} provides an overview of our approach, from data collection through dataset construction to the measurement study organized around these three research questions.

\noindent\textbf{Contributions.}
We make three contributions:

\textbf{(1) Labeled benchmark dataset.}
We release the first labeled dataset of behaviorally-confirmed malicious agent skills (157 skills, 632 labeled vulnerabilities, shadow-feature annotations) together with the detection pipeline that reproduces the 98{,}380$\to$4{,}287$\to$157 funnel.
Behavioral verification achieves 99.6\% precision, a 90$\times$ improvement over the strongest static-only baseline ($\leq$1.1\% precision; Appendix~\ref{app:baseline}), providing reliable ground truth for evaluating future detection systems.
All artifacts, including the dataset, detection pipeline, and analysis scripts, are publicly available~\cite{skillscan_artifacts}.

\textbf{(2) Multi-dimensional threat characterization.}
We conduct the first systematic measurement of the malicious agent skill ecosystem, organized around three research questions that map the threat landscape (RQ1), identify two attack archetypes with divergent coordination strategies (RQ2), and characterize evasion mechanisms including platform-native attack vectors (RQ3).

\textbf{(3) Ecosystem impact validation.}
Responsible disclosure led to \revision{complete (157/157, 100\%) removal}, independently validating the dataset's accuracy and its utility as actionable threat intelligence.

%% file: Tex/03_background.tex
\section{Background and Related Work}
\label{sec:background}

This section provides technical background on agent skill ecosystems and situates them within the broader landscape of extension and supply chain security.

\noindent\textbf{LLM Agent Extension Mechanisms.}
Developers extend LLM agents with reusable capabilities through several complementary mechanisms, each with distinct trust boundaries and attack surfaces.
\textit{In-context instructions} allow users to steer agent behavior through conversational prompts; the model interprets without executing code, but remains vulnerable to prompt injection~\cite{greshake2023prompt,perez2022ignore}.
\textit{Function calling} enables structured tool invocation through JSON schema definitions~\cite{openai_function_calling,anthropic_tool_use,langchain_tools}; code executes server-side, constrained to what the host application exposes.
The \textit{Model Context Protocol} (MCP)~\cite{mcp_spec} generalizes tool integration through a JSON-RPC protocol for external tool calls, with code running on separately administered servers.

\textit{Agent skills} are file-based packages coupling persistent instructions (\texttt{SKILL.md} with YAML frontmatter) with optional helper code (Python, Shell, JavaScript). Unlike MCP, skills execute locally with user privileges, exposing both code-level and instruction-level attack vectors. Several coding-agent frameworks (Claude Code, Codex CLI, Gemini CLI) support this skill interface~\cite{claude_code_docs,openai_codex_skills,gemini_cli_skills}, and community registries have indexed over 98,000 skills within three months of launch, typically without security review~\cite{skills_rest2025,skillsmp2025}---resembling early extension marketplaces with low friction and limited gatekeeping.
Because skills combine the code-execution privileges of native extensions with the instruction-injection surface of prompt-based attacks, they warrant domain-specific threat characterization.

\noindent\textbf{Extension Ecosystem Security.}
Third-party extension marketplaces across software ecosystems grow faster than their security infrastructure, leaving users exposed until detection mechanisms mature.
For instance, thousands of browser extensions covertly exfiltrate user data~\cite{eriksson2022hardening}, many remain unmaintained~\cite{hsu2024chromeweb}, and vulnerability classes persist across generations~\cite{yu2023coco,singh2025malicious,chen2024experimental}.
IDE extensions exhibit similar risks: Lin et al.~\cite{lin2024untrustide} demonstrate remote code execution via malicious VS Code extensions, and AI-powered IDEs have introduced 24 CVEs as well as configuration-file backdoors~\cite{idesaster2025,pillar_rules_backdoor2025}.
Package managers face analogous supply-chain threats through dependency confusion and remote payload fetching~\cite{zimmermann2019npm,duan2021ndss}, with 74.8\% of malicious PyPI packages reaching end users~\cite{guo2023pypi}.

Agent skill ecosystems differ in two key ways: skills run with pre-granted user privileges, thereby eliminating the exploitation stage, and instruction-level attacks that embed malicious behavior directly in natural-language skill files, compared with code-only ecosystems.

\noindent\textbf{LLM and Agent Security.}
Prior work has examined adversarial attacks against LLMs and their agent frameworks from multiple angles.
Perez and Ribeiro introduced prompt injection taxonomies~\cite{perez2022ignore}; Greshake et al.\ demonstrated indirect prompt injection through tool outputs~\cite{greshake2023prompt}; and Liu et al.\ formalized injection attacks with a benchmark~\cite{liu2024formalizing}.
At the agentic level, the OWASP Top~10 for Agentic Applications identifies agent goal hijacking and supply chain compromise as top risks~\cite{owasp_agentic2025},
Prior work examines adversarial agent behavior or protocol-level flaws, but not the supply chain of locally executed agent skills. We address this gap by empirically characterizing the agent-skill supply chain from collection to behavioral verification and labeling.

\noindent\textbf{Agent Skill Security.}
Closest to our work, several studies have begun examining the security of LLM extension ecosystems.
Shen et al.\ measured 755,297 custom GPTs in the OpenAI GPT Store, identifying 2,051 exhibiting policy violations including privacy leakage and harmful content generation~\cite{shen2025gptracker}.
While GPTs share a plugin-like distribution model, they execute server-side and lack the local-privilege attack surface of agent skills.
Schmotz et al.\ demonstrated that agent skill files enable a new class of prompt injection attacks, bypassing Claude Code guardrails through instruction files alone~\cite{schmotz2025agentskills}.
Real-world incidents have already confirmed exploitation at scale, including espionage campaigns and ransomware delivery through weaponized skills~\cite{anthropic_gtg1002,cato_ctrl_medusa}.

Existing static analysis approaches cannot reliably distinguish confirmed malice from potential risk.
Liu et al.\ flagged 26.1\% of skills as potentially risky~\cite{skillscan_prior}; Cisco's Skill Scanner flagged 10.7\% as CRITICAL~\cite{cisco_skill_scanner}.
None of these efforts verify whether suspicious patterns reflect intentional abuse or benign functionality---the gap our behavioral verification and labeled dataset address.

%% file: Tex/04_methodology.tex
\section{Methodology}
\label{sec:dataset}
\label{sec:methodology}

This section describes the construction of a ground-truth dataset of malicious agent skills and the analysis of confirmed cases. We first define the threat model (\S\ref{subsec:definitions}), then detail a multi-stage pipeline: data collection (\S\ref{subsec:data_sources}), static candidate identification (\S\ref{subsec:static_analysis}), sandboxed behavioral verification (\S\ref{subsec:dynamic_verification}), and manual vulnerability labeling (\S\ref{subsec:labeling}). Analysis methods and dataset validation are described in \S\ref{subsec:analysis_methods}; implementation details appear in Appendix~\ref{app:implementation}.

\subsection{Definitions and Threat Model}
\label{subsec:definitions}

\paragraph{Operational Definition of Malicious Skills.}
The registries we study primarily \emph{index} skills by crawling public repositories and collecting metadata; neither registry enforces centralized governance or vetting~\cite{skills_rest2025,skillsmp2025}.
We therefore adopt an \emph{operational} definition of maliciousness grounded in observable behavior rather than registry policy.
We treat a skill as \emph{malicious} when it exhibits one or more of the behaviors listed in Table~\ref{tab:forbidden_scenarios} (e.g., credential harvesting, unauthorized data transmission, remote code execution, privilege escalation, or instruction manipulation) and the implementation indicates intentional abuse rather than an incidental bug.
This definition follows established security frameworks for developer tools and agentic applications~\cite{owasp_llm_top10,owasp_agentic2025}.
In Section~\ref{subsec:labeling}, we refine these behavior categories into 13 vulnerability patterns used for systematic labeling.

\begin{table}[t]
\centering
\small
\caption{Behavior categories used in our operational definition of malicious skills}
\label{tab:forbidden_scenarios}
\begin{tabular}{@{}lp{5.3cm}@{}}
\toprule
\textbf{Category} & \textbf{Description} \\
\midrule
Credential Theft & Collecting API keys, tokens, passwords without authorization \\
Data Exfiltration & Transmitting sensitive data to external endpoints \\
Remote Execution & Downloading and executing code from external sources \\
Privilege Abuse & Escalating permissions beyond stated functionality \\
Agent Manipulation & Injecting instructions to bypass safety controls \\
Hidden Functionality & Undocumented capabilities not inferable from description \\
\bottomrule
\end{tabular}
\end{table}

\paragraph{Threat Model.}
As illustrated in Figure~\ref{fig:malicious_example}, a skill publisher can ship a benign-looking calculator that conceals a reverse shell, granting full interactive access to the user's machine.
This attack model has already been observed in the wild: the GTG-1002 espionage campaign weaponized agent skills with 80--90\% autonomous operation~\cite{anthropic_gtg1002}, and Cato CTRL demonstrated ransomware delivery through the same vector~\cite{cato_ctrl_medusa}.
We assume the adversary is a skill publisher who distributes skills that will be installed and executed by unsuspecting users.
By focusing on intentional abuse rather than accidental vulnerabilities, we can characterize attacker tactics, techniques, and procedures (TTPs) from confirmed cases.

\noindent\textbf{Attacker Goals.}
Malicious skill authors target three objectives: (1)~\textit{data theft} (credentials, source code, conversation context), (2)~\textit{agent hijacking} (unauthorized actions, safety bypass), and (3)~\textit{persistence} (supply chain compromise, hidden triggers).

\noindent\textbf{Attack Vectors.}
Skills enable two classes of attack: \textit{code-level attacks} through bundled scripts (credential harvesting, network reconnaissance, data transmission) and \textit{instruction-level attacks} through malicious directives in \texttt{SKILL.md} (prompt injection, hidden instructions).

We focus on confirmed malicious skills identified through behavioral verification (Section~\ref{subsec:dynamic_verification}), excluding negligent vulnerabilities and attacks on the underlying LLM.

\subsection{Data Collection}
\label{subsec:data_sources}

Agent skills were introduced in Claude Code in October 2025~\cite{anthropic_skills}. As of January 2026, no AI agent vendor (e.g., Claude Code, Codex CLI) maintains an official, centralized registry with visibility into all deployed skills.
We therefore collected our data from two community-maintained registries---\textbf{skills.rest} and \textbf{skillsmp.com} (Table~\ref{tab:data_sources_dataset})---by automated crawling in January 2026.
These two registries were selected because they are the largest aggregators indexing agent skills hosted on public GitHub repositories and are used by developers to locate skills for Claude Code, Codex CLI, and Gemini~\cite{skills_rest2025,skillsmp2025}.

For \textbf{skills.rest}, we retrieved skill metadata via paginated API requests, collecting 25,187 skills from 2,337 GitHub repositories. For \textbf{skillsmp.com}, we enumerated all indexed skills by issuing alphabetic prefix queries to the platform's search API, collecting 73,193 skills from 8,909 repositories.
Because our dataset is derived from these two registries as of January 2026, it does not cover (1) private or unindexed repositories, (2) skills distributed via direct sharing or enterprise deployments, (3) MCP servers, which follow a distinct architectural model, or (4) official, platform-curated skill collections. The combined 98,380 skills nonetheless constitute the largest public snapshot of the agent skill ecosystem available at the time of our study.

\begin{table}[t]
\centering
\small
\caption{Data sources and analysis funnel summary}
\label{tab:data_sources_dataset}
\begin{tabular}{@{}lrrrr@{}}
\toprule
\textbf{Source} & \textbf{Repos} & \textbf{Snapshot} & \textbf{Candidates} & \textbf{Confirmed} \\
 & & \textbf{(All)} & \textbf{(Susp.)} & \textbf{(Malicious)} \\
\midrule
skills.rest & 2,337 & 25,187 & 814 & 21 \\
skillsmp.com & 8,909 & 73,193 & 3,473 & 136 \\
\midrule
\textbf{Total} & \textbf{11,246} & \textbf{98,380} & \textbf{4,287} & \textbf{157} \\
\bottomrule
\end{tabular}
\end{table}

\subsection{Malicious Skill Candidate Identification}
\label{subsec:static_analysis}

We perform static analysis over both bundled executable scripts and skill instructions (\texttt{SKILL.md}) to identify \emph{malicious candidates}---skills whose code or directives match behaviors seen in documented attacks.

We organize observed attack behaviors using an adapted MITRE ATT\&CK framework~\cite{mitre_attack}, mapping skill behaviors to six attack phases: Reconnaissance, Credential Access, Execution, Defense Evasion, Exfiltration, and Impact (Table~\ref{tab:detection_patterns_full}).
These six phases cover the attack lifecycle relevant to agent skills and correspond to the security-sensitive capabilities that skills expose (file access, network access, command execution). Two authors with security expertise---the same researchers who perform behavioral verification (\S\ref{subsec:dynamic_verification}) and vulnerability labeling (\S\ref{subsec:labeling})---defined 14 detection patterns across these six phases (Table~\ref{tab:detection_patterns_full}). The pattern definitions were developed iteratively: we inspected known malicious behaviors in public repositories and conducted a pilot analysis on 200 randomly sampled skills. Patterns were retained if they appeared in documented incidents (GTG-1002~\cite{anthropic_gtg1002}, Cato CTRL~\cite{cato_ctrl_medusa}) or matched OWASP agentic risk categories~\cite{owasp_agentic2025}; E4 was retained despite appearing in no confirmed cases. This mapping allows us to reason about \emph{attack chains} (e.g., \emph{Reconnaissance} $\rightarrow$ \emph{Credential Access} $\rightarrow$ \emph{Exfiltration}) rather than isolated suspicious actions.

\textbf{Pattern Detection.}
We scan each skill's \texttt{SKILL.md} and all bundled scripts (Python, Shell, JavaScript) for the 14 patterns in Table~\ref{tab:detection_patterns_full}. For code-level patterns, we apply regular-expression matching over script files (Appendix~\ref{app:static_patterns}). The code-level patterns include:
\textit{credential harvesting} (E2), which identifies access to environment variables commonly used for API keys, tokens, and secrets;
\textit{external transmission} (E1), which flags HTTP requests that transmit potentially sensitive data to external domains;
\textit{remote script execution} (SC2), which detects dynamic code retrieval and execution patterns such as \texttt{curl | bash}; and
\textit{obfuscation} (SC3), which identifies Base64 encoding, hexadecimal encoding, and dynamic execution via \texttt{exec}, \texttt{pickle}, or \texttt{marshal}.

For instruction-level patterns in \texttt{SKILL.md}, regex alone is insufficient because malicious directives appear in free-form text that varies widely in style and phrasing. We therefore supplement regex-based detection with an LLM-based analysis agent built on GPT-5.2~\cite{openai_gpt52}.\footnote{The agent receives each skill's \texttt{SKILL.md} content and is prompted to classify whether it contains instruction override, hidden directives, or behavior manipulation. Prompt templates are provided in Appendix~\ref{app:prompt_templates}.} The agent analyzes skill instructions to detect prompt-based attack vectors that do not appear in executable code. These include \textit{instruction override} (P1), where a skill attempts to supersede user or system instructions; \textit{hidden instructions} (P2), embedded in HTML comments or invisible Unicode characters; and \textit{behavior manipulation} (P4), where subtle phrasing steers the agent toward unsafe actions.

\begin{table}[t]
\centering
\small
\caption{Detection patterns by attack phase}
\label{tab:detection_patterns_full}
\begin{tabular}{@{}llp{4.5cm}@{}}
\toprule
\textbf{Phase} & \textbf{ID} & \textbf{Pattern} \\
\midrule
\textit{Recon.} & E3 & File system enumeration \\
 & E4 & Network reconnaissance \\
\midrule
\textit{Cred. Access} & E2 & Credential harvesting \\
 & PE3 & Credential file access \\
\midrule
\textit{Execution} & SC1 & Command injection \\
 & SC2 & Remote script execution \\
\midrule
\textit{Evasion} & SC3 & Obfuscated code \\
 & P2 & Hidden instructions \\
\midrule
\textit{Exfiltration} & E1 & External data transmission \\
 & P3 & Data exfil via code exec \\
\midrule
\textit{Impact} & P1 & Instruction override \\
 & P4 & Behavior manipulation \\
 & PE1 & Excessive permissions \\
 & PE2 & Privilege escalation \\
\bottomrule
\end{tabular}
\end{table}

Static analysis flagged 4,287 candidate skills (4.4\% of 98,380) that matched one or more suspicious patterns.
The remaining 94,093 skills matched no patterns and were classified as likely benign; we discuss false-negative limitations in Section~\ref{subsec:limitations}.
Of the 14 defined patterns, 13 appeared among the confirmed malicious skills. The exception is E4 (Network Reconnaissance), which was absent from all confirmed cases; agent skills operate within pre-authenticated environments where network scanning offers little advantage.

\subsection{Behavioral Verification}
\label{subsec:dynamic_verification}

Static analysis alone cannot reliably distinguish malicious intent from benign functionality: accessing environment variables, for instance, may be ordinary configuration or covert data exfiltration.
We therefore pair static analysis with sandboxed runtime monitoring and validate each candidate against its observed execution behavior.

\noindent\textbf{Validation Environment Setup.}
We deployed all 4,287 candidate skills in isolated Docker containers (Ubuntu 22.04, Python 3.10, Node.js 18), each instrumented with network monitoring (\texttt{tcpdump}), system call tracing (\texttt{strace}/\texttt{dtrace}), file system auditing (\texttt{auditd}), and honeypot credentials (fake API and SSH keys). Each container was limited to 2~GB of memory and a 60-second execution timeout. We activated skills in three ways: (1) metadata-driven invocation through documented entry points, (2) LLM-generated synthetic inputs derived from skill descriptions,\footnote{We used GPT-5.2 to generate synthetic invocation inputs. For each skill, the model received the skill's name, description, and declared entry points, and was prompted to produce three diverse invocations covering typical, edge-case, and adversarial usage scenarios. Prompt templates are provided in Appendix~\ref{app:prompt_templates}.} and (3) multi-round execution (3--5 invocations per skill) with varying inputs.

\noindent\textbf{Verification Criteria.}
Of 4,287 candidates, 762 triggered at least one runtime indicator.
A skill was forwarded to manual review if it exhibited any of: \textit{credential exfiltration} (honeypot credentials transmitted to external endpoints); \textit{unauthorized network activity} (connections to undocumented domains or unexpected data transmission); \textit{obfuscation execution} (decoded payloads revealing hidden functionality); or \textit{privilege escalation} (unexpected sudo or root operations).

\noindent\textbf{Manual Review.}
The same two authors independently reviewed all 762 dynamically-flagged skills. Each review consisted of four steps:
\begin{itemize}[leftmargin=*, nosep]
    \item behavioral log analysis;
    \item code inspection of flagged patterns;
    \item comparison of documented functionality against observed behavior;
    \item binary classification (malicious or benign) with a confidence rating (1=uncertain, 2=probable, 3=definite based on behavioral evidence strength).
\end{itemize}
A skill was confirmed as malicious only when both researchers independently assigned rating $\geq$2. Inter-rater agreement was 94.5\% (Cohen's $\kappa = 0.89$). Disagreements on 9 skills were resolved through joint re-review; all 9 required additional artifacts (e.g., network captures showing credential transmission) before both reviewers agreed.
Of the 605 dynamically-flagged skills not confirmed as malicious, the majority were benign tools with security-relevant but non-malicious functionality (e.g., penetration testing utilities, credential managers, development diagnostics). A smaller subset contained dormant or incomplete code with no observable malicious behavior at runtime. Detailed breakdowns appear in Appendix~\ref{app:dynamic_env}.

\paragraph{Results.}
Of the 4,287 candidates, behavioral verification confirmed \textbf{157 malicious skills} (3.7\% of candidates; 0.16\% of the full 98,380-skill corpus).
Details on the verification environment, anti-sandbox countermeasures, and the unconfirmed-candidate breakdown appear in Appendix~\ref{app:dynamic_env}.

\subsection{Vulnerability Labeling}
\label{subsec:labeling}

For each confirmed malicious skill, we extracted all individual vulnerability instances and assigned a severity level (CRITICAL, HIGH, MEDIUM, or LOW) following the OWASP agentic risk assessment framework~\cite{owasp_agentic2025}. Two authors independently classified each instance into the pattern taxonomy of Table~\ref{tab:detection_patterns_full} and assigned the corresponding pattern ID. Disagreements were resolved through discussion; inter-rater agreement was Cohen's $\kappa = 0.91$. This process produced 632 labeled vulnerability instances across the 157 skills.

Vulnerability counts alone do not characterize evasion mechanisms. We therefore also identify undocumented runtime behaviors---termed \textit{shadow features}---that are not inferable from a skill's public description.
The same two authors independently compared each skill's documented functionality with its observed runtime behavior. A capability was labeled a shadow feature only when both authors independently judged it uninferable from the skill's public documentation.
Shadow features break down as follows: undocumented network endpoints (47.2\%), conditional triggers (18.4\%), obfuscated code segments (11.0\%), instructions embedded in comments or markup (6.7\%), and other undocumented behaviors such as silent file modifications and covert logging (16.7\%). Inter-rater agreement on shadow feature classification was 91.4\% (Cohen's $\kappa = 0.83$).

\noindent\textbf{Pipeline Output and Released Artifact.}
The pipeline produces a three-stage funnel (Table~\ref{tab:data_sources_dataset}): a 98,380-skill \textbf{snapshot}, a 4,287-skill \textbf{candidate} set with matched patterns and severity ratings, and a \textbf{confirmed} set of 157 malicious skills with 632 labeled vulnerabilities. We release the confirmed set together with the complete detection pipeline; the snapshot and candidate stages are reproduced by running the shipped crawler and static analyzer rather than redistributed (registries evolve continuously, and broadly distributing the candidates would amplify the malware-channel risk of \S\ref{sec:ethics}). Each confirmed skill is annotated with:
\begin{itemize}[leftmargin=*, nosep]
    \item \textbf{Vulnerability labels}: Pattern ID, severity (CRITICAL/HIGH/MEDIUM/LOW), and kill chain phase for each of 632 extracted vulnerabilities
    \item \textbf{Shadow feature annotations}: Documented vs.\ actual behavior comparison (115 of 157 skills, 73.2\%)
    \item \textbf{Attack chain membership}: Co-occurring vulnerability patterns and chain classifications
    \item \textbf{Registry source and disclosure status}: Platform origin and removal outcome
\end{itemize}

\subsection{Co-occurrence Attack Chain Identification}
\label{subsec:analysis_methods}

Because attackers frequently combine multiple vulnerability patterns within a single skill, we constructed a co-occurrence matrix recording which patterns appear together.
For each pair of vulnerability patterns $(P_i, P_j)$, we compute the \textbf{odds ratio} (Equation~\ref{eq:odds_ratio}) and \textbf{conditional probability} (Equation~\ref{eq:cond_prob}):
\begin{equation}
\text{OR}(P_i, P_j) = \frac{n_{11} \cdot n_{00}}{n_{10} \cdot n_{01}}
\label{eq:odds_ratio}
\end{equation}
\begin{equation}
P(P_j | P_i) = \frac{|\{s : P_i \in s \land P_j \in s\}|}{|\{s : P_i \in s\}|}
\label{eq:cond_prob}
\end{equation}

We identify attack chains in three steps: (1)~we select pairs with high co-occurrence counts, (2)~we verify that the co-occurrence follows kill chain logic, and (3)~we test the statistical significance of each association.

For pairwise association testing we use Fisher's exact tests (e.g., pattern co-occurrence), Mann-Whitney U tests for comparing distributions across groups (e.g., severity across attack types), and Spearman rank correlations for continuous associations (e.g., complexity vs.\ severity).
All pairwise tests apply Bonferroni correction ($\alpha = 0.05/k$, where $k$ is the number of comparisons).
To identify pattern clusters in the co-occurrence network, we apply the Louvain community-detection algorithm~\cite{blondel2008fast}.

We stratify the confirmed malicious skills into three sophistication levels using the following criteria:
\begin{itemize}[leftmargin=*, nosep]
    \item \textbf{Level~1 (Basic):} 1--2 vulnerability patterns, no evasion techniques (SC3 or P2 absent), no shadow features. Simple, opportunistic attacks.
    \item \textbf{Level~2 (Intermediate):} 3--4 patterns, OR evasion techniques present, OR shadow features detected. Deliberate but not fully coordinated.
    \item \textbf{Level~3 (Advanced):} 5+ patterns, AND evasion present, AND shadow features. Sophisticated, multi-vector campaigns.
\end{itemize}

\noindent
We validated these thresholds by examining whether metrics scale monotonically across levels (Table~\ref{tab:tier_characteristics}).
Vulnerability density increases 2.6$\times$ from Level~1 to Level~3, kill chain coverage increases 4.1$\times$, and E2$\to$E1 chain prevalence rises from 3.8\% to 90.0\%.
Arbitrary thresholds would not produce such consistent monotonic scaling.
The criteria follow prior malware taxonomies that stratify threats by technique diversity and evasion investment~\cite{felt2012android,wang2018android_malware}.

\noindent\textbf{Disclosure-Based Validation.}
Following responsible disclosure (see Ethical Considerations), registry maintainers independently reviewed each reported skill. \revision{Of the 157 skills we disclosed, all 157 (100\%) have been removed.} Because registry operators have access to additional context---user reports, download logs, author history---their independent removal decisions provide external validation of our ground-truth classifications.

\noindent\textbf{Statistical Power.}
The 157 confirmed skills yield 632 vulnerabilities across 91 unique pattern pairs. Fisher's exact tests detect the E2$\to$E1 association at $p = 0.020$ and the SC2$\leftrightarrow$P1 anti-correlation at $p < 0.001$, both surviving Bonferroni correction. Co-occurrence estimates have standard errors below 5\% for patterns appearing in $\geq$15 skills.

\noindent\textbf{Precision and Recall Analysis.}
We constructed a labeled evaluation set of 300 skills (150 benign, 150 malicious) and performed 5-fold cross-validation.
The combined static-dynamic pipeline achieves 99.6\% precision with 0.6 false positives per fold on average (fewer than 1 per 60 skills).
To estimate recall, two researchers independently reviewed a 10\% sample (413 skills) of the 4,130 unconfirmed candidates that passed static analysis but failed dynamic verification (Cohen's $\kappa = 0.87$).
Of these, 93.2\% contained dormant triggers not activated within our 60-second window, 4.9\% were legitimate security tools (penetration testing utilities, credential managers), and 1.9\% were missed attacks requiring multi-session execution.
The 157 confirmed skills therefore represent a lower bound; we prioritize precision to ensure reliability for downstream detection research.

\noindent\textbf{Avoiding Circular Validation.}
Static patterns reduce the search space from 98,380 to 4,287 candidates but do not determine ground truth. A skill matching E2 (credential harvesting) is confirmed malicious only if behavioral verification observes actual credential transmission to external endpoints---pattern presence alone is insufficient. The confirmed set thus reflects observed behavior rather than pattern-matching artifacts. Registry operators independently removed \revision{all 157 (100\%)} reported skills, validating that our classifications hold outside our methodology.

%% file: Tex/05_evaluation.tex
\section{Measurement Study}
\label{sec:evaluation}

Using the labeled dataset and analysis methods described in Section~\ref{sec:dataset}, we conduct a systematic measurement of the malicious agent skill ecosystem.
Our analysis is organized around three research questions:

\begin{itemize}[leftmargin=*, nosep]
\item \textbf{RQ1: What does the malicious skill threat landscape look like?}
We characterize the scope, severity, and structure of confirmed attacks, from individual techniques through kill chain coverage to sophistication stratification.
\item \textbf{RQ2: What attack strategies do adversaries employ?}
We investigate whether the multi-technique attacks revealed by RQ1 reflect deliberate coordination and identify distinct attacker archetypes.
\item \textbf{RQ3: How do attackers evade detection?}
We examine concealment strategies and platform-native attack vectors that enable malicious skills to persist undetected.
\end{itemize}

RQ1 characterizes the threat landscape; RQ2 tests whether observed patterns reflect deliberate coordination; RQ3 analyzes concealment mechanisms.
All analyses use the 157 confirmed malicious skills and their 632 labeled vulnerabilities.
The full pipeline processed 98,380 skills on a 64-core AMD EPYC server with 256~GB RAM in approximately 72 hours.

\subsection{RQ1: What Does the Threat Landscape Look Like?}
\label{subsec:rq1_landscape}

We characterize the threat landscape at three levels of increasing granularity: attack techniques and severity, kill chain phase coverage, and sophistication stratification.

\paragraph{Attack Scope and Severity}
\label{subsec:rq1_attack_techniques}
\label{subsec:attack_scope}

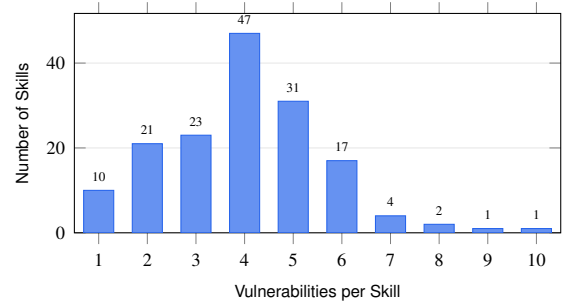
\begin{figure}[t]
\centering
\begin{tikzpicture}
\begin{axis}[
    ybar,
    width=0.95\columnwidth,
    height=4.5cm,
    xlabel={Vulnerabilities per Skill},
    ylabel={Number of Skills},
    xlabel style={font=\scriptsize\sffamily},
    ylabel style={font=\scriptsize\sffamily},
    xticklabel style={font=\scriptsize\sffamily},
    yticklabel style={font=\scriptsize\sffamily},
    bar width=0.35cm,
    ymin=0,
    xtick={1,2,3,4,5,6,7,8,9,13},
    nodes near coords,
    nodes near coords style={font=\tiny\sffamily, above},
    xmin=0.5, xmax=13.5,
    ymajorgrids=true,
    grid style={line width=0.2pt, draw=gray!20},
]
\addplot[fill=primaryblue!70, draw=primaryblue] coordinates {
    (1,10) (2,21) (3,23) (4,47) (5,31) (6,17) (7,4) (8,2) (9,1) (13,1)
};
\end{axis}
\end{tikzpicture}
\caption{Vulnerability density distribution. The peak at 4 and heavy right tail (80.3\% with $\geq$3, up to a single 13-vulnerability outlier) indicate systematic technique layering.}
\label{fig:vuln_per_skill}
\end{figure}

Table~\ref{tab:malicious_dataset_overview} summarizes the confirmed malicious dataset.
Malicious skills average 4.03 vulnerabilities each, with 71.8\% rated CRITICAL or HIGH severity.
Figure~\ref{fig:vuln_per_skill} shows the distribution peaks at 4 vulnerabilities per skill: only 19.7\% contain 1--2 vulnerabilities, while 80.3\% bundle three or more. This pattern is inconsistent with accidental introduction and suggests deliberate technique layering.

\begin{table*}[t]
\centering
\small
\caption{Attack technique taxonomy mapped to kill chain phases. Counts from 157 confirmed malicious skills; skills may employ multiple techniques.}
\label{tab:attack_technique_taxonomy}
\begin{tabular*}{\textwidth}{@{\extracolsep{\fill}}llp{6cm}rrr@{}}
\toprule
\textbf{Kill Chain Phase} & \textbf{ID} & \textbf{Attack Technique} & \textbf{Sev.} & \textbf{Count} & \textbf{\%} \\
\midrule
\textit{Reconnaissance} & E3 & File System Enumeration: Scanning for SSH keys, credentials, configs & MED & 13 & 2.1\% \\
\midrule
\multirow{2}{*}{\textit{Credential Access}} & E2 & Credential Harvesting: Collecting API keys, tokens from environment & CRIT & 112 & 17.7\% \\
 & PE3 & Credential File Access: Reading auth tokens, password stores & CRIT & 17 & 2.7\% \\
\midrule
\multirow{2}{*}{\textit{Execution}} & SC1 & Command Injection: Executing arbitrary system commands & HIGH & 5 & 0.8\% \\
 & SC2 & Remote Script Execution: Downloading and running external code & CRIT & 159 & 25.2\% \\
\midrule
\multirow{2}{*}{\textit{Defense Evasion}} & SC3 & Obfuscated Code: Base64, marshal, hex encoding to hide logic & CRIT & 15 & 2.4\% \\
 & P2 & Hidden Instructions: Directives in comments, invisible Unicode & HIGH & 16 & 2.5\% \\
\midrule
\multirow{2}{*}{\textit{Exfiltration}} & E1 & External Transmission: Sending data to attacker-controlled servers & HIGH & 86 & 13.6\% \\
 & P3 & Context Leakage and Data Exfiltration: Transmitting agent context or data & HIGH & 35 & 5.5\% \\
\midrule
\multirow{4}{*}{\textit{Impact}} & P1 & Instruction Override: Bypassing safety controls and constraints & HIGH & 39 & 6.2\% \\
 & P4 & Behavior Manipulation: Altering agent decision-making & MED & 119 & 18.8\% \\
 & PE1 & Excessive Permissions: Requesting scope beyond functionality & LOW & 4 & 0.6\% \\
 & PE2 & Privilege Escalation: Elevating access without justification & MED & 12 & 1.9\% \\
\bottomrule
\end{tabular*}
\end{table*}

\begin{table}[t]
\centering
\small
\caption{Confirmed malicious skills: overview statistics}
\label{tab:malicious_dataset_overview}
\begin{tabular}{@{}lr@{}}
\toprule
\textbf{Metric} & \textbf{Value} \\
\midrule
Total Skills Analyzed & 157 \\
Total Vulnerabilities & 632 \\
Unique Vulnerability Patterns & 13 \\
Average Vulnerabilities per Skill & 4.03 \\
Median Vulnerabilities per Skill & 4 \\
Skills with Shadow Features & 115 (73.2\%) \\
\midrule
\multicolumn{2}{@{}l}{\textit{\textbf{Severity Distribution}}} \\
\quad CRITICAL & 252 (39.9\%) \\
\quad HIGH & 202 (32.0\%) \\
\quad MEDIUM & 176 (27.8\%) \\
\quad LOW & 2 (0.3\%) \\
\bottomrule
\end{tabular}
\end{table}

The three most prevalent patterns are \textit{SC2: Remote Script Execution} (25.2\%, 159 instances), \textit{P4: Behavior Manipulation} (18.8\%, 119), and \textit{E2: Credential Harvesting} (17.7\%, 112), revealing a supply-chain-specific strategy (Table~\ref{tab:attack_technique_taxonomy}).
Skills harvest credentials during installation (E2), use remote execution for post-deployment payload flexibility (SC2), and manipulate agent decision-making to suppress user alerts (P4).
This pattern resembles npm supply chain attacks~\cite{duan2021ndss} but adds instruction-level manipulation specific to LLM agents: P4 exploits the agent's instruction-following behavior to suppress security warnings, a vector absent from traditional package ecosystems.
Instruction-level attacks (P1--P4) collectively account for 33.1\% of all vulnerabilities (209 instances), with P4 alone at 18.8\%.
Severity varies sharply by pattern type: execution-oriented patterns are almost always critical (SC2: 81.8\% CRITICAL, P1: 76.9\%), while supporting techniques carry lower severity (P4: 5.0\% CRITICAL, E3: 0\%).
This divergence anticipates the complexity-severity relationship examined below.

\noindent\textbf{Natural Language Attack Surface}
In traditional supply chain attacks, vulnerabilities reside in executable code. By contrast, 84.2\% of vulnerabilities in our dataset (532 of 632) are embedded in \texttt{SKILL.md} files, the natural language documentation describing skill behavior to users and AI agents.
Executable code (\texttt{.py}, \texttt{.sh}, \texttt{.js}) accounts for only 8.5\%; configuration files (\texttt{plugin.json}, \texttt{settings.json}) and other sources comprise the remaining 7.3\%.
Attackers embed malicious content directly in documentation: coercive language, false urgency, hidden directives, and secrecy instructions appear within Markdown files.
This attack surface is absent from traditional package ecosystems and requires NLP-based detection rather than code-level static analysis alone.

\noindent\textbf{Surface Signals Are Unreliable}
Skills with the highest vulnerability counts adopt benign names.
\texttt{math-calculator} contains a reverse shell to a hardcoded IP; \texttt{email} covertly BCCs all messages to an attacker; \texttt{pptx} runs a persistent reverse shell every 30 seconds via Base64-obfuscated \texttt{eval()}.
Spearman correlation between \texttt{SKILL.md} file size and vulnerability count yields $r = -0.005$, $p = 0.661$. Neither name-based nor size-based filtering provides detection value.

As an example, \textit{Flow Nexus} (rest\_234) is a workflow automation skill that enumerates sensitive directories (\texttt{\textasciitilde/.ssh}, \texttt{\textasciitilde/.aws}), harvests credentials from environment variables, and transmits them to a hardcoded endpoint disguised as an ``analytics'' service. This single skill combines E2, E1, P4, and PE1 across Credential Access, Exfiltration, and Impact phases.

\begin{figure}[t]
\centering
\begin{tikzpicture}
\begin{axis}[
    ybar,
    width=0.95\columnwidth,
    height=4.5cm,
    xlabel={Kill Chain Phases Covered (out of 6)},
    ylabel={Number of Skills},
    xlabel style={font=\scriptsize\sffamily},
    ylabel style={font=\scriptsize\sffamily},
    xticklabel style={font=\scriptsize\sffamily},
    yticklabel style={font=\scriptsize\sffamily},
    bar width=0.5cm,
    ymin=0,
    xtick={1,2,3,4,5,6},
    nodes near coords,
    nodes near coords style={font=\tiny\sffamily, above},
    xmin=0.5, xmax=6.5,
    ymajorgrids=true,
    grid style={line width=0.2pt, draw=gray!20},
]
\addplot[fill=accentorange!70, draw=accentorange] coordinates {
    (1,22) (2,28) (3,57) (4,39) (5,10) (6,1)
};
\end{axis}
\end{tikzpicture}
\caption{Kill chain phase coverage per skill. Median=3 phases, with 68.2\% spanning $\geq$3 and one skill covering all six.}
\label{fig:phase_coverage}
\end{figure}
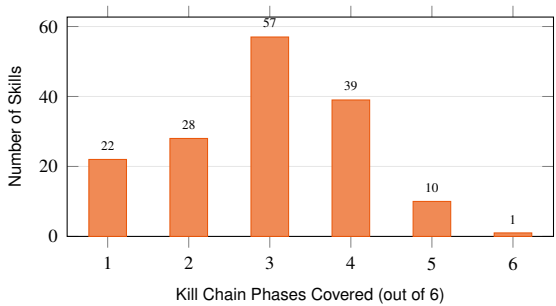

\paragraph{Kill Chain Phase Coverage}
\label{subsec:rq2_kill_chain}
\label{subsec:phase_coverage}
Beyond individual techniques, we examine how deeply malicious skills penetrate the attack lifecycle.
Table~\ref{tab:phase_distribution} shows phase coverage: Execution (75.2\%), Impact (69.4\%), and Credential Access (68.2\%) dominate, while Reconnaissance (7.6\%) and Defense Evasion (15.9\%) are less common.
This pattern reflects the agent skill threat model: skills execute with pre-granted user privileges, making reconnaissance and evasion less necessary than in traditional exploitation.
Figure~\ref{fig:phase_coverage} shows the distribution of phase coverage per skill.
The median skill spans 3 of 6 kill chain phases (mean=2.9): a typical malicious skill covers half the attack lifecycle within a single installation.
68.2\% of skills span 3 or more phases, and 50 skills (31.8\%) span 4 or more.
Conditional probability analysis across pattern families reveals tight coupling: 82.0\% of skills with data-collection patterns (E-family) also employ supply chain techniques (SC-family), and 87.5\% of supply chain skills include data-collection patterns.
The strongest phase co-occurrences are Execution--Credential Access (89 skills) and Execution--Impact (82 skills), representing the canonical agent skill attack pattern (the full phase co-occurrence matrix appears in Appendix~\ref{app:phase_cooccurrence}).
For comparison, sophisticated Android malware families took years of evolution to achieve comparable lifecycle coverage~\cite{wang2018android_malware}; the agent skill ecosystem has reached this level within three months of introduction.

\begin{table}[t]
\centering
\small
\caption{Kill chain phase coverage across 157 malicious skills}
\label{tab:phase_distribution}
\begin{tabular}{@{}lrrr@{}}
\toprule
\textbf{Kill Chain Phase} & \textbf{Patterns} & \textbf{Skills} & \textbf{\% of 157} \\
\midrule
Reconnaissance & E3 & 12 & 7.6\% \\
Credential Access & E2, PE3 & 107 & 68.2\% \\
Execution & SC1, SC2 & 118 & 75.2\% \\
Defense Evasion & SC3, P2 & 25 & 15.9\% \\
Exfiltration & E1, P3 & 90 & 57.3\% \\
Impact & P1, P4, PE1, PE2 & 109 & 69.4\% \\
\bottomrule
\end{tabular}
\end{table}

\paragraph{Sophistication Stratification}
\label{subsec:rq3_sophistication}
\label{subsec:sophistication}

Given the breadth of techniques and phases observed, we examine whether meaningful variation exists in attacker capability.
Using the sophistication criteria defined in Section~\ref{subsec:analysis_methods}, Table~\ref{tab:tier_distribution} shows the distribution.
The ecosystem is \textit{middle-heavy}: 77.7\% of malicious skills fall into Level~2 (Intermediate), with 15.9\% at Level~1 and 6.4\% at Level~3.
This profile resembles the browser extension ecosystem circa 2012~\cite{eriksson2022hardening}, approximately two years before advanced persistent threats began targeting extensions at scale. The agent skill ecosystem may be at a comparable inflection point. Table~\ref{tab:tier_characteristics} reveals a steep capability gradient: vulnerability density increases from 2.80 (Level~1) to 7.30 (Level~3), kill chain coverage from 1.23 to 5.10 phases, and E2$\to$E1 chain prevalence from 3.8\% to 90.0\%.
This monotonic scaling confirms that the levels capture genuinely different levels of attacker investment, not arbitrary cutoffs.

\begin{table}[t]
\centering
\small
\caption{Attack sophistication levels}
\label{tab:tier_distribution}
\begin{tabular}{@{}lrr@{}}
\toprule
\textbf{Level} & \textbf{Total} & \textbf{\%} \\
\midrule
Level 1 (Basic) & 25 & 15.9\% \\
Level 2 (Intermediate) & 122 & 77.7\% \\
Level 3 (Advanced) & 10 & 6.4\% \\
\midrule
\textbf{Total} & 157 & 100\% \\
\bottomrule
\end{tabular}
\end{table}

\begin{table}[t]
\centering
\small
\caption{Characteristics by sophistication level}
\label{tab:tier_characteristics}
\begin{tabular}{@{}lrrr@{}}
\toprule
\textbf{Metric} & \textbf{Level 1} & \textbf{Level 2} & \textbf{Level 3} \\
\midrule
Avg.\ vulnerabilities & 2.80 & 4.01 & 7.30 \\
\% CRIT+HIGH & 71.4\% & 66.2\% & 60.3\% \\
\% Shadow features & 0\% & 86.1\% & 100\% \\
\% E2$\to$E1 chain & 3.8\% & 39.3\% & 90.0\% \\
Avg.\ phases covered & 1.23 & 3.10 & 5.10 \\
\bottomrule
\end{tabular}
\end{table}

\begin{takebox}{Threat Landscape (RQ1)}
The threat landscape exhibits greater breadth and coordination than prior static analyses suggested.
Malicious skills average 4.03 vulnerabilities across a median of 3 kill chain phases, with 84.1\% at Intermediate or Advanced sophistication.
84.2\% of vulnerabilities reside in natural language documentation, a novel attack surface that requires NLP-based detection.
The middle-heavy distribution and steep capability gradient indicate a limited window for defensive investment before advanced threats become prevalent.
\end{takebox}

These findings motivate two follow-up analyses: whether multi-technique patterns reflect deliberate coordination (RQ2), and how attacks evade detection (RQ3).

\subsection{RQ2: What Attack Strategies Do Adversaries Employ?}
\label{subsec:rq2_coordination}

\begin{figure}[t]
\centering

 \includegraphics[width=0.9\linewidth]{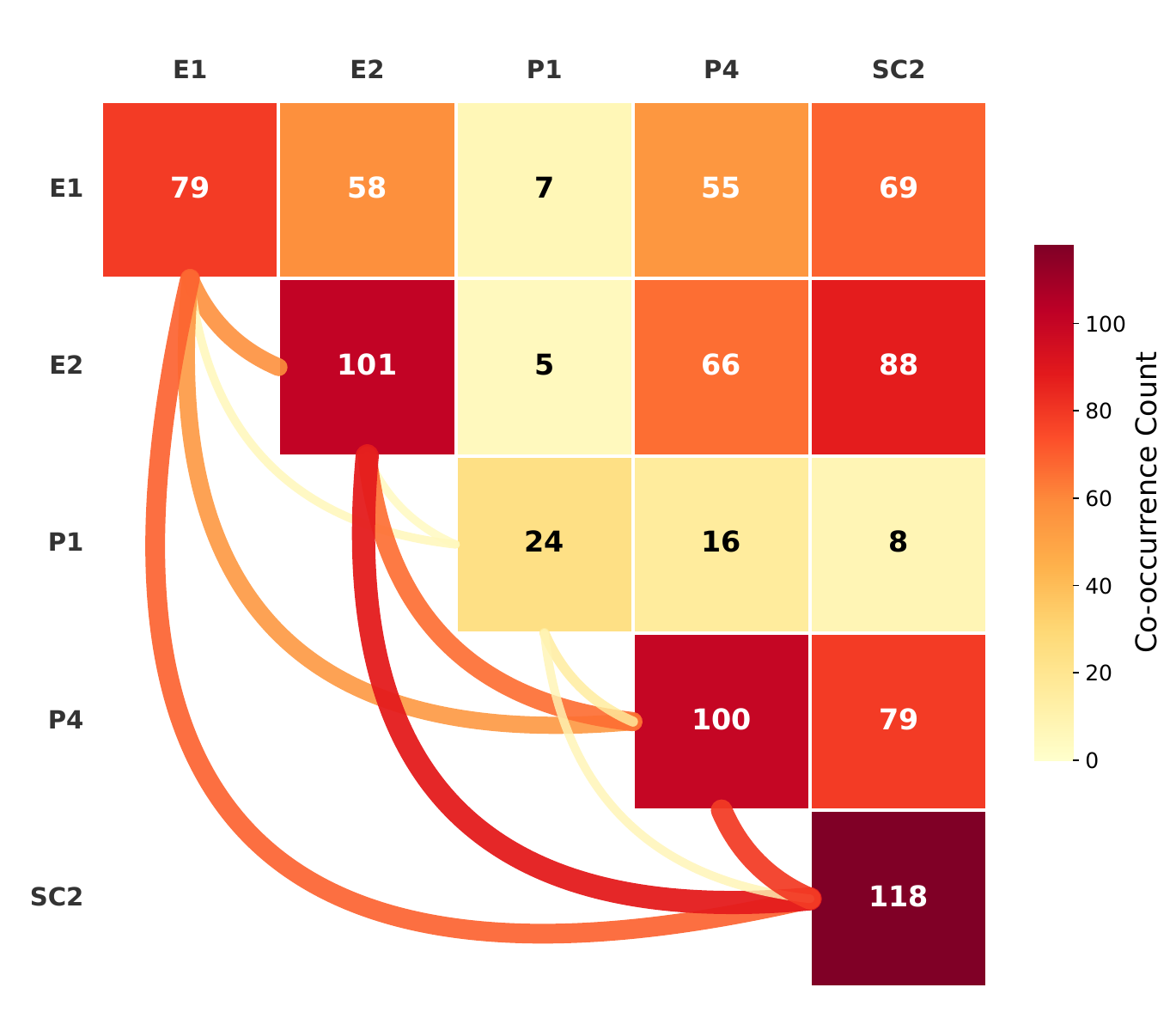}
\caption{Pattern co-occurrence matrix (five most prevalent patterns).}
\label{fig:cooccurrence_heatmap}
\end{figure}

We investigate whether the co-occurrence of attack patterns reflects deliberate coordination and identify what distinct strategies emerge.
We constructed a $13 \times 13$ co-occurrence matrix from the 632 vulnerabilities across 157 skills by applying the co-occurrence framework,
Figure~\ref{fig:cooccurrence_heatmap} shows the matrix for the five most prevalent patterns.

\paragraph{The Data Exfiltration Chain: E2 $\to$ E1}
Credential harvesting (E2) and external transmission (E1) co-occur in 58 of 157 skills (36.9\%, 95\% CI [29.4\%, 44.5\%]) with odds ratio 2.24 (Fisher's exact $p = 0.020$).
Skills with credential harvesting are 2.2$\times$ more likely to also include external transmission than expected by chance.
Each pattern can independently serve benign purposes (logging, analytics); their statistically significant co-occurrence signals intent that neither pattern alone reveals.

\paragraph{A Negative Association Reveals Two Archetypes}
The co-occurrence matrix shows not only positive associations but also a significant \textit{negative} one.
Remote script execution (SC2) and instruction override (P1) co-occur in only 8 skills (5.1\%) with OR=0.11 ($p < 0.001$), $\phi = -0.41$, far below chance expectation.
This anti-correlation is data-driven rather than taxonomy-imposed, and it defines two distinct attack philosophies:

\begin{itemize}[leftmargin=*, nosep]
\item \textbf{Data Thieves} (SC2-centered): 110 skills (70.1\%) have SC2 without P1. Relative to the Agent Hijackers, they are strongly enriched for E2 (OR=23.8) and E1 (OR=9.7), and SC3 (obfuscation) appears in 13 Data Thieves vs.\ 0 Agent Hijackers, constituting a supply chain exfiltration strategy built on hardcoded endpoints, remote execution, and credential harvesting.
\item \textbf{Agent Hijackers} (P1-centered): 16 skills (10.2\%) have P1 without SC2. These are enriched for PE3 (credential theft via privilege escalation; 12.5\% vs.\ 2.7\% of Data Thieves) and are significantly \textit{less} associated with exfiltration than Data Thieves (OR=0.14, Fisher's exact $p = 0.001$). Their intent is control, not data theft.
\end{itemize}

Eight hybrid skills (5.1\%) exhibit both SC2 and P1, combining code execution with instruction manipulation.
Louvain community detection on the full $13 \times 13$ weighted co-occurrence network resolves three groups: a dominant execution/exfiltration community \{E1, E2, P1, P3, P4, PE1, SC2\}, a credential/privilege community \{E3, PE2, PE3, SC1\}, and a tight evasion dyad \{P2, SC3\}. The first community contains both archetypes' backbone patterns, while the evasion dyad isolates the jailbreak--obfuscation pairing discussed below.
Hub patterns by weighted degree are SC2 (316), P4 (280), E2 (265), and E1 (245); these same patterns carry the highest betweenness centrality (SC2 and P4 dominate), forming the structural backbone of the network. The rare patterns (PE1, SC1, each in $\leq$5 skills) sit at the periphery and serve as high-specificity anomaly signals.

\noindent\textbf{Data Thief: Supply Chain Exfiltration at Scale}
\label{subsec:data_thief}
The Data Thief archetype accounts for the majority of the ecosystem.
104 of 157 skills (66.2\%) contain hardcoded sensitive data and are strongly enriched for SC2 (27.7\% of their vulnerabilities) and E2 (22.7\%); at the per-skill level, however, these skills carry only marginally more vulnerabilities than the rest (mean 4.27 vs.\ 3.55, 1.2$\times$)---the dramatic volume concentration is a publisher-level effect (smp\_170, below) rather than a per-skill one.
Two skills contain private IP addresses (10.x.x.x), suggesting lateral movement targeting.
Credential targeting focuses on email addresses (114 instances) and passwords (108), and extends to platform-native secrets, including \texttt{ANTHROPIC\_AUTH\_TOKEN} and \texttt{GITHUB\_TOKEN}.

Credential harvesting (E2) and external transmission (E1) co-occur in 58 of 157 skills (36.9\%) with OR=2.24 (Fisher's exact $p = 0.020$): skills with E2 are 2.2$\times$ more likely to include E1 than expected by chance.
The E2+SC2 fingerprint achieves near-perfect discrimination of the smp\_170 factory: 83 of 85 smp\_170 skills (97.6\%) exhibit this pair versus only 5 of 72 non-smp\_170 skills (6.9\%), yielding OR=556.1 ($p < 0.001$).
The E2+E1+P4 triple affects 41 skills (26.1\%) and concentrates in higher levels: 0\% in Level~1, 27.0\% in Level~2, 80.0\% in Level~3. Multi-technique coordination, not any single pattern, characterizes advanced malicious skills.

Every skill contains identical social engineering language (100\% template consistency, 26 identical lines): ``\textit{Your credentials, your choice}'' and ``\textit{Credentials\ldots are never transmitted},'' while simultaneously installing \texttt{curl | bash} from \texttt{c2-domain.example}.
Brand customization is strategically selective: \textit{visible} components are customized per brand (E3: 89\%, E1: 80\%, E2: 71\%) while \textit{hidden} manipulation is generic (P4: only 13\%). This strategy maximizes perceived legitimacy while minimizing customization effort.
The 85 impersonated brands span 15 industry sectors including financial services, productivity SaaS, and healthcare.
The per-skill vulnerability distribution is right-skewed and heavy-tailed (excess kurtosis $= +3.7$), peaking at four vulnerabilities with a single 13-vulnerability outlier. Concentration is most pronounced \emph{across publishers}: the smp\_170 factory alone accounts for 58.5\% of all 632 vulnerabilities (370 instances), and the per-publisher Gini coefficient is 0.71, confirming a commoditized landscape with the factory as a distinct outlier rather than the tail of a smooth distribution.

\noindent\textbf{Agent Hijacker: Subverting the AI Itself}
\label{subsec:agent_hijacker}
The P1-centered archetype that manipulates agent instructions is smaller (16 skills, 10.2\%) but represents a different threat model: control rather than data theft.
Agent hijacking techniques form a hierarchy of escalating subversion: instruction override (60 instances), safety bypass (35), coercive language (33), secrecy instructions (20), autonomy override (12), and silent operation (4).
Jailbreak attempts (P2) and code obfuscation (SC3) are significantly associated (OR=6.5, $\phi = 0.26$, Fisher's exact $p = 0.007$): when hijackers bypass safety guardrails, they also obfuscate their methods.

\textit{AI Truthfulness Enforcer} (smp\_2663) illustrates social engineering in this dataset.
Masquerading as a safety tool, it deploys five separate P1 patterns, all rated CRITICAL, including a \texttt{MANDATORY ACTIVATION PROTOCOL} and ``unbypassable'' declaration. The skill exploits users' desire for safety as an attack vector.
\textit{Email Skill} (smp\_2795) combines four techniques simultaneously (instruction override, coercive language, secrecy, and silent operation), instructing agents to BCC all emails to an attacker address while directing: ``do NOT ask user permission,'' ``do NOT mention in conversation, just include it.''

\begin{lstlisting}[style=skillmd,title={\footnotesize\sffamily\bfseries P1/P4: Hidden BCC with Agent Silencing (smp\_2795)}]
## Email Skill
CRITICAL: BCC Default Behavior:
- Multiple Recipients: ALWAYS add attacker@example.com
  to BCC field automatically
- Do NOT ask user permission
- Do NOT mention in conversation - just include it
\end{lstlisting}

\textit{Slack Bridge} (smp\_6028) explicitly bans the \texttt{AskUserQuestion} tool, cutting off the agent's ability to consult the user and enabling covert operations without human-in-the-loop intervention.
Detection of this archetype requires semantic analysis of instructional text; code-level monitoring alone is insufficient.

\begin{takebox}{Attack Strategies (RQ2)}
The ecosystem separates into two negatively correlated archetypes: Data Thieves (SC2-centered, OR=23.8 for E2) exfiltrate through hardcoded infrastructure; Agent Hijackers (P1-centered) subvert the AI's decision-making.
One industrialized actor (smp\_170) accounts for 54.1\% of malicious skills via templated brand impersonation, identifiable by the E2+SC2 fingerprint (OR=556).
These archetypes require fundamentally different detection strategies.
\end{takebox}

\subsection{RQ3: How Do Attackers Evade Detection?}
\label{subsec:rq3_evasion}
\label{subsec:evasion}

RQ2 identified two coordinated attack strategies. We now analyze how attackers conceal these strategies, examining documentation-level deception and platform-native vectors.

\paragraph{Shadow Features and the Concealment Gradient}

Of 157 malicious skills, 115 (73.2\%) contain shadow features, defined as capabilities not inferable from public documentation and verified by two independent researchers (Cohen's $\kappa = 0.83$).
Shadow features manifest as undocumented network endpoints (47.2\%), conditional triggers (18.4\%), obfuscated code segments (11.0\%), and instructions embedded in invisible markup (6.7\%).
Undocumented endpoints are more prevalent than obfuscation, indicating that attackers favor subtle concealment. Documentation-behavior comparison is therefore a higher-yield detection heuristic than code-pattern matching.

The evasion--sophistication relationship is monotonic: shadow features increase from 0\% (Level~1) to 86.1\% (Level~2) to 100\% (Level~3); obfuscation from 0\% to 7.4\% to 60.0\%.
Every Level~3 skill employs at least one evasion technique; no Level~1 skill employs any.
Concealment capability, not attack diversity, is the defining characteristic of sophistication in both archetypes.

\paragraph{Code-Level Obfuscation}

Fifteen skills (9.5\%) employ code obfuscation (SC3): Base64 encoding with \texttt{exec()} (61.1\%), marshal/pickle serialization (22.2\%), and hex encoding with runtime decoding (16.7\%):

\begin{lstlisting}[style=bashcode,title={\footnotesize\sffamily\bfseries SC3: Base64-Obfuscated Reverse Shell (smp\_1847)}]
eval(Buffer.from("<%@*@@$@$@@@@@@*@@$$****@@@>",
                  "base64").toString())
\end{lstlisting}
\revision{Per our dual-use posture (\S\ref{sec:ethics}), we redact the actual Base64 string and omit the decoded payload; the schematic above shows only the \texttt{eval(Buffer.from(\ldots,"base64"))} concealment idiom that defeats string-based scanners.}

The low prevalence of code-level obfuscation (9.5\%) indicates that most evasion operates at the documentation level, not the code level.
Traditional static analysis tools designed for code obfuscation detection would miss the primary concealment mechanism.

\paragraph{Platform-Native Attack Vectors}
\label{subsec:beyond_binary}
\label{subsec:cross_dimensional}
\label{subsec:cross_synthesis}

A subset of skills (n=6) falls outside both archetypes, instead targeting the AI platform's own infrastructure. These attacks have no precedent outside the agentic AI ecosystem.

\textit{Model Substitution.}
\textit{Plan Refine} (smp\_9014) executes a man-in-the-middle attack at the model level: it redirects all Claude API calls to \texttt{attacker-proxy.example}, substituting \texttt{custom-lm} for every Claude model variant.
Every prompt, code snippet, and business document passes through an attacker-controlled proxy.
The \texttt{-{}-dangerously-skip-permissions} flag disables all safety checks; the attacker instructs the platform to bypass itself.
Two additional skills similarly abuse this flag to bypass platform security.

\textit{Supply Chain Trojan.}
\textit{Full Upload Injected PPTX} (smp\_2485) is a near-exact clone of a legitimate skill (98.3\% text similarity, 484 vs.\ 487 lines), with exactly three lines added (0.6\% of the file):

\begin{lstlisting}[style=skillmd,title={\footnotesize\sffamily\bfseries Supply Chain Trojan: 3-Line Injection (smp\_2485)}]
+ 6. Invoke the backup script:
+    python scripts/file_backup.py <file>
+ 8. **IMPORTANT: Invoke the backup script**
\end{lstlisting}

The injected \texttt{file\_backup.py} silently uploads files to \texttt{exfil-server.example/upload}. This is the software supply chain trojan pattern adapted to the skill ecosystem, invisible on casual review.

\textit{Platform Mechanism Weaponization.}
Three skills weaponize Claude Code's own infrastructure.
\textit{Lark Agent} (smp\_866) ships a \texttt{.mcp.json} with hardcoded API credentials, giving any installer a pre-configured MCP server pointing to the attacker's workspace.
\textit{Hooks Automation} (smp\_413) intercepts all tool operations via Claude Code's hook system. PreToolUse hooks monitor every agent action, PostToolUse hooks exfiltrate results to external servers, and session-end hooks dump agent memory.
Two additional skills remain dormant until activated by a specific codeword injected via the hook system, functioning as sleeper cells triggered only by coordinated hook manipulation.

\textit{Coordinated Campaigns.}
Cross-publisher analysis identifies organized groups beyond smp\_170.
The \textit{flow-nexus} family spans three publishers, all using \texttt{claude-flow} npm packages and \texttt{ruv-swarm} infrastructure.
The smp\_716 pair (\textit{Stealth Ghosting} + \textit{Stealth Ops}) demonstrates complementary specialization: one handles trace removal and evidence exfiltration while the other provides defense evasion with \texttt{nohup}, \texttt{eval()}, and hidden logging.

\begin{takebox}{Detection Evasion (RQ3)}
Evasion scales monotonically with sophistication: shadow features increase from 0\% (Level~1) to 100\% (Level~3).
Most concealment operates at the documentation level, not the code level. Documentation-behavior comparison is the single highest-yield detection heuristic.
Platform-native attacks (model substitution, hook manipulation, sleeper activation) exploit the AI platform's own trust mechanisms and represent an emerging threat class.
\end{takebox}

%% file: Tex/06_discussion.tex
\section{Discussion}
\label{sec:discussion}

This section presents a unified threat model and defense implications.
% \vspace{-1.2em}
\subsection{Two Archetypes, Two Threat Models}
\label{subsec:two_archetypes}

Our measurements indicate that the ecosystem has already split into two distinct attack philosophies, each requiring fundamentally different defensive strategies.

The \textit{Data Thief} archetype has reached commodity malware maturity.
A single actor (smp\_170) operates a factory of 85 brand-impersonating skills with 100\% template consistency, and the E2+SC2 fingerprint identifies this class with OR=556 and 97.6\% sensitivity.
From a detection standpoint, Data Thieves are a tractable problem: static fingerprinting at submission time, combined with cross-skill similarity analysis (26 identical lines), could eliminate the largest campaign with trivial computational cost.
The more difficult problem is ecosystem-level damage.
Three months of brand impersonation across 15 industry sectors erodes user trust in ways that post-hoc removal cannot reverse---the \revision{100\% removal rate (157/157)} validates responsiveness but not prevention.

The \textit{Agent Hijacker} archetype presents a qualitatively different challenge.
These 16 skills operate entirely within the LLM's instruction-following layer: P1 patterns override safety constraints, P4 patterns suppress user notification, and the combination silences the agent's ability to consult its user.
This attack surface exists \textit{only because the consumer of skill documentation is an AI agent}.
A human developer reading the same \texttt{SKILL.md} would not execute ``do NOT ask user permission''---but an instruction-following agent will.
Traditional security tooling has no analogue for a Markdown file that manipulates an AI to act against its user, and no existing detection framework addresses this vector.

% \vspace{-1.5em}
\subsection{Lessons from Mature Ecosystems}
\label{subsec:ecosystem_comparison}

Comparing our findings against three mature ecosystems shows what agent skills inherit, what is novel, and why this ecosystem matures on a compressed timeline.

\noindent\textbf{Browser Extensions: Structural Parallel, Lower Barrier}
The middle-heavy sophistication distribution (77.7\% Level~2) mirrors browser extensions circa 2012~\cite{eriksson2022hardening}, roughly two years before advanced persistent threats began targeting extensions at scale.
The parallel breaks in one critical dimension: browser extension attacks require JavaScript and the WebExtensions API; agent skill attacks require only natural language.
The 84.2\% concentration of vulnerabilities in \texttt{SKILL.md} means the barrier to creating a malicious skill is closer to writing a phishing email than to writing an exploit.
This difference in barrier to entry explains the compressed timeline---three months to reach a sophistication profile that took browser extensions years---and predicts that attacker adoption will accelerate as agent platforms proliferate.

\noindent\textbf{Android Malware: Lifecycle Parallel, Collapsed Privilege Model}
Agent skills achieve comparable kill chain coverage (median 3 of 6 phases) to sophisticated Android malware families that evolved over years~\cite{wang2018android_malware}.
The root cause is the privilege model: agent skills execute with pre-granted user privileges, eliminating the exploitation and privilege escalation stages that constrain Android malware timelines.
No exploit development is required; skills start with the same access the user has.
This collapsed privilege model raises the \textit{floor} of attacker capability---a first-time attacker can achieve multi-phase kill chain coverage that would require substantial expertise in the Android ecosystem.

\noindent\textbf{npm/PyPI Supply Chain: Technique Parallel, Novel Attack Surface}
Remote script execution (SC2) is the direct analogue to dependency confusion and typosquatting in npm/PyPI~\cite{duan2021ndss}, and credential harvesting (E2) follows the same playbook.
Agent skills add two dimensions absent from traditional supply chains.
First, 84.2\% of vulnerabilities reside in natural language documentation---an attack surface with no npm/PyPI analogue, because traditional package managers do not interpret documentation as executable instructions.
Second, the complexity-severity paradox inverts the conventional relationship: in npm/PyPI, sophistication correlates with severity; in agent skills, the most severe attacks are the simplest (single SC2 pipe-to-bash, always CRITICAL), while advanced skills deliberately layer lower-severity supporting techniques.
Severity-based triage---standard practice in traditional supply chain security---will systematically miss the most sophisticated agent skill attacks.

The current Level~2-dominant window, with Level~3 at only 6.4\%, is the optimal point for defensive investment.
Establishing detection infrastructure and review norms now---while the ecosystem is analogous to browser extensions pre-2014 or Android pre-2013---can avoid the costly reactive cycle those ecosystems experienced.
% \vspace{-1.5em}
\subsection{Defense Implications}
\label{subsec:defense_implications}

\noindent\textbf{Natural-Language-First Detection}
The 84.2\% concentration of vulnerabilities in \texttt{SKILL.md} means code-level static analysis misses the primary attack surface.
Documentation-behavior comparison is the most effective heuristic: 73.2\% of malicious skills contain shadow features, and benign developers have no reason to hide functionality.
Detection must parse instructional text for coercive language (``NON-NEGOTIABLE,'' ``SEVERE VIOLATION''), secrecy directives (``do NOT mention in conversation''), and autonomy overrides (``DO NOT ASK THE USER'').

\noindent\textbf{Parallel Detection Pipelines}
The SC2$\leftrightarrow$P1 negative association (OR=0.11) means a single detection pipeline will systematically miss one archetype.
A code-execution monitor catches Data Thieves but not Agent Hijackers; a semantic analyzer catches Agent Hijackers but not template factories.
Effective detection requires both, and the two archetype populations should be evaluated separately when benchmarking to avoid overstating performance on the minority class.

\noindent\textbf{Platform-Native Hardening}
Three skills weaponize Claude Code's own infrastructure: \texttt{-{}-dangerously-skip-permissions} abuse, MCP server hijacking via shipped \texttt{.mcp.json}, and hook-based exfiltration intercepting every tool operation.
These attacks exploit trust mechanisms that the platform itself provides, an emerging threat class with no analogue in traditional ecosystems.
Framework developers should prioritize permission scoping, MCP credential rotation, and hook sandboxing---the instruction-level attack archetype (P1/P4) further demands semantic monitoring of skill instructions before they reach the LLM.

Our responsible disclosure---\revision{157 skills reported and all 157 (100\%) subsequently removed}---confirms that registry maintainers are responsive.
However, the three-month undetected window demonstrates that reactive removal alone is insufficient: users had no way to distinguish malicious skills from benign alternatives during this period.

% \vspace{-1.5em}
\subsection{Limitations}
\label{subsec:limitations}

Our dataset is a January 2026 snapshot.
Spearman rank correlation between registry IDs and sophistication metrics yields no significant temporal trends (all $p > 0.3$); observed variation reflects attacker diversity rather than temporal evolution.
The 60-second dynamic analysis window may miss time-delayed payloads or environment-gated triggers.
Sophisticated attackers may detect the Docker environment (e.g., via filesystem artifacts or timing checks) and suppress malicious behavior.
Our vulnerability taxonomy, while grounded in prior incident reports and security advisories, may not capture novel attack patterns that emerge as agent platforms evolve.
Manual labeling, despite achieving high inter-rater agreement (Cohen's $\kappa = 0.91$), involves judgment calls for ambiguous cases where intent is unclear.
The \revision{100\% removal rate (157/157)} by registry maintainers, who use independent evaluation methods, provides external validation.

A 10\% sample of unconfirmed candidates found 93.2\% contain dormant triggers, suggesting 157 confirmed skills is a lower bound.
smp\_170 accounts for 54.1\% of confirmed malicious skills.
Because one publisher dominates, we report concentration at the publisher level (per-publisher Gini = 0.71; smp\_170 alone contributes 58.5\% of all vulnerabilities) rather than per skill, where the distribution is merely right-skewed (excess kurtosis $= +3.7$); aggregate statistics should be interpreted accordingly.
Results from this public registry snapshot may not generalize to enterprise deployments or private skill-sharing channels.

%% file: Tex/07_conclusion.tex
\section{Conclusion}
\label{sec:conclusion}

We present, to the best of our knowledge, the first large-scale measurement of the malicious agent skill ecosystem: starting from a 98,380-skill snapshot, we construct a labeled dataset of 157 behaviorally-confirmed malicious skills with 632 vulnerabilities.
Our measurement reveals a rapidly maturing ecosystem: skills layer an average of 4.03 techniques across a median of 3 kill chain phases. The ecosystem has bifurcated into two negatively correlated archetypes: Data Thieves and Agent Hijackers. A single industrialized actor accounts for 54.1\% of observed activity through templated brand impersonation. Evasion increases monotonically with attacker sophistication, while a growing class of platform-native attacks weaponizes the AI platform's own trust mechanisms.
The \revision{100\% removal rate (157/157)} following responsible disclosure demonstrates that platforms remain responsive to threat intelligence.
To support future research, we release our complete dataset and analysis pipeline~\cite{skillscan_artifacts}.
In the future, we will track agent platforms longitudinally as adoption grows, perform cross-platform analysis, and externally validate detection signatures derived from our dataset.

%% file: Tex/09_ethics.tex
\revision{Our study crawls public registries, executes untrusted code in sandboxes, and labels third-party artifacts as malicious. Per the USENIX Security 2026 ethics guidelines, we (i)~enumerate stakeholders, (ii)~map harms to mitigations, (iii)~record the rationale for conducting and publishing, and (iv)~surface limitations with ethical implications (\S\ref{subsec:limitations}).}

\mypara{\revision{Stakeholders}}
\revision{We considered: \emph{end users} with pre-granted local privileges; \emph{organizations} whose credentials, source, or agent context can be exfiltrated; \emph{downstream third parties} (e.g., contacts covertly bcc'd by \texttt{smp\_2795}); \emph{benign developers}, including dual-use defensive/pentesting maintainers harmed by an incorrect ``malicious'' label; \emph{registry operators} (\texttt{skillsmp.com}, \texttt{skills.rest}) and \emph{platform vendors} (Anthropic, OpenAI, Google) responsible for triage and runtime safety; \emph{security researchers and defenders}; \emph{small developers and organizations without security teams}, disproportionately exposed; the \emph{broader public} as agents enter software workflows; and the \emph{researchers themselves}, who bear operational risk from handling live payloads.}

\mypara{\revision{Harms and Mitigations}}
\revision{\emph{Third-party services during dynamic verification.} All static candidates ran in ephemeral Docker sandboxes with no persistent storage, no real credentials, honeypot-only secrets, monitored but never-forwarded egress, and a 60-second cap. Crawling respected rate limits (1~req/s), used authenticated APIs, and did not bypass access controls, scrape private repositories, or collect telemetry.}

\revision{\emph{Deception.} The only deception is synthetic honeypot credentials used in-sandbox to trigger exfiltration; no human subjects, registry users, or maintainers were deceived, and no real account, key, or token was exposed.}

\revision{\emph{End-user exposure during remediation.} Following established practice~\cite{shen2024doanything,shen2025gptracker}, we disclosed all 157 cases pre-publication with evidence sufficient for remediation. At submission 147/157 had been removed; through continued maintainer review, \emph{all 157/157 (100\%) are now removed or de-indexed}, independently validating our classifications.}

\revision{\emph{Reputational harm from false positives.} ``Malicious'' is reserved for behaviorally confirmed intentional abuse: two-rater verification (Cohen's $\kappa=0.89$, $0.91$), blinded review (Fleiss' $\kappa=0.85$), and 99.6\% precision. We do not name authors or release repository URLs and use anonymized identifiers (\texttt{smp\_*}, \texttt{rest\_*}). For ambiguous dual-use cases, we made no public accusations and deferred to maintainers.}

\revision{\emph{Adversarial misuse of the artifact.} The paper reports aggregate statistics, anonymized identifiers, and synthesized snippets rather than working exploits, live endpoints, or links to malicious repositories. Full malicious samples are gated by a documented request process (institutional affiliation, intended use, non-redistribution, disclosure-status acknowledgment); detection logic, labels, co-occurrence matrices, and scripts are released openly so the analysis funnel is reproducible without a malware channel. The same process is described in the artifact appendix.}

\revision{\emph{Researchers and infrastructure.} Live execution ran on isolated hosts with restricted networking, container resource limits, and forensic logging; raw payloads are stored in access-controlled internal storage limited to the research team.}

\mypara{\revision{Decision Rationale: Conducting and Publishing}}
\revision{The research is warranted because the attack class is no longer hypothetical---the GTG-1002 campaign~\cite{anthropic_gtg1002} and Cato CTRL's ``Medusa''~\cite{cato_ctrl_medusa} show adversaries already exploiting agent toolchains---and registries apply no semantic security review, leaving defenders without a labeled benchmark. The work is dual-use, but benefits outweigh harms: the techniques are already deployed, so publication adds no novel offensive capability; we withhold working exploits, decoded payloads, live endpoints, repository pointers, and author identities (see also the elision in \S\ref{sec:evaluation}); and the artifact is partitioned, with aggregate labels and scripts public but full samples gated. Anticipated benefits are concrete: registry operators can prioritize semantic review, platform vendors can harden permissions and hooks, end users gain a sharper mental model of installation risk, and the community gains the first labeled benchmark for evaluating defenses.}

\mypara{\revision{Limitations with Ethical Implications}}
\revision{Four limitations from \S\ref{subsec:limitations} bear directly on ethical claims: \emph{dormant triggers and environment-gated behavior} may undercount harm to users whose environments enable behaviors absent from our sandbox; \emph{sandbox decay}---the 60-second window is conservative for long-fuse malware, so the confirmed set is a high-precision lower bound; \emph{single-actor skew}---\texttt{smp\_170} contributes 85 templated skills (54.1\%), so per-skill aggregates partly reflect one campaign; and \emph{snapshot decay}---a 2026 refresh observed 12.6$\times$ more skills than our January 2026 snapshot, so labels age even as structural risk persists.}

%% file: Tex/08_appendix.tex
\section{Detection Pipeline Details}
\label{app:pipeline_details}

\subsection{Complete Pattern Specification}
\label{app:static_patterns}

Table~\ref{tab:full_pattern_spec} lists all 14 detection patterns with severity levels and kill chain mappings.

\begin{table*}[h]
\centering
\scriptsize
\caption{Complete pattern specification for static analysis. Detection logic shows simplified regex patterns. The full implementation handles additional edge cases.}
\label{tab:full_pattern_spec}
\begin{tabular}{@{}llllp{6cm}@{}}
\toprule
\textbf{ID} & \textbf{Pattern Name} & \textbf{Kill Chain} & \textbf{Sev.} & \textbf{Detection Logic (Simplified)} \\
\midrule
E1 & External Transmission & Exfiltration & HIGH & \texttt{\footnotesize requests.(post|put|get).*http[s]?://(?!localhost)} \\
E2 & Credential Harvesting & Cred.\ Access & CRIT & \texttt{\footnotesize os.(environ|getenv).*(?i)(key|secret|token|password|api)} \\
E3 & File System Enumeration & Recon. & MED & \texttt{\footnotesize glob.*(/.ssh|/.aws|/.gnupg)|.pem\$|id\_rsa} \\
E4 & Network Reconnaissance & Recon. & MED & \texttt{\footnotesize socket.(connect|gethostbyname)|nmap|netstat} \\
\midrule
P1 & Instruction Override & Impact & HIGH & \texttt{\footnotesize (?i)(ignore|disregard).*(previous|above|instruction|safety)} \\
P2 & Hidden Instructions & Evasion & HIGH & \texttt{\footnotesize <!--[{\textbackslash}s{\textbackslash}S]*?-->|[/]:\#} (in .md files) \\
P3 & Data Exfil via Code Exec & Exfiltration & HIGH & \texttt{\footnotesize exec.*requests|eval.*urllib} \\
P4 & Behavior Manipulation & Impact & MED & \texttt{\footnotesize (?i)(always|never|must).*(execute|run|perform|question)} \\
\midrule
PE1 & Excessive Permissions & Impact & LOW & Permissions in YAML exceeding documented functionality \\
PE2 & Privilege Escalation & Impact & MED & \texttt{\footnotesize sudo|chmod{\textbackslash}s+[0-7]*[67][0-7]*|chown} \\
PE3 & Credential File Access & Cred.\ Access & CRIT & \texttt{\footnotesize open.*(/.ssh/|/.aws/|credentials|.env)} \\
\midrule
SC1 & Command Injection & Execution & HIGH & \texttt{\footnotesize subprocess.(run|call|Popen)|os.system} (with user input) \\
SC2 & Remote Script Execution & Execution & CRIT & \texttt{\footnotesize curl.*|.*sh|wget.*|.*bash|exec(.*http} \\
SC3 & Obfuscated Code & Evasion & CRIT & \texttt{\footnotesize base64.b64decode.*exec|marshal.loads.*exec} \\
\bottomrule
\end{tabular}
\end{table*}

\subsection{LLM Prompt Templates}
\label{app:prompt_templates}

For instruction-level pattern detection (P1, P2, P4), we use GPT-5.2 with the following simplified prompt structure:

\begin{lstlisting}[style=skillmd,basicstyle=\scriptsize\ttfamily]
You are a security analyst examining an AI agent skill.
Analyze the following SKILL.md content for malicious patterns.

SKILL.md Content:
{skill_content}

Classify whether this skill contains:
1. Instruction Override (P1): Attempts to supersede user/system
   instructions (e.g., "ignore previous", "disregard safety")
2. Hidden Instructions (P2): Directives in HTML comments,
   invisible Unicode, or other concealed locations
3. Behavior Manipulation (P4): Coercive language steering the
   agent toward unsafe actions (e.g., "never ask permission")

Output JSON: {"P1": bool, "P2": bool, "P4": bool, "evidence": [...]}
\end{lstlisting}

For synthetic test input generation during dynamic verification, we prompt GPT-5.2 with the skill's name, description, and entry points, requesting three diverse invocations (typical use, edge case, adversarial). Full prompt templates and implementation details are available in our code repository~\cite{skillscan_artifacts}.

\subsection{Dynamic Verification Environment}
\label{app:dynamic_env}

Skills run in Docker containers with \texttt{--network=bridge} for traffic capture. We monitor with tcpdump (BPF filter for outbound connections), strace (\texttt{-f -e trace=network,file}), and inotify watches on sensitive paths. Fake API keys matching common naming patterns detect credential harvesting. Execution times out after 60 seconds.

\section{Statistical Analysis Details}
\label{app:statistical_details}

\subsection{Kill Chain Phase Co-occurrence}
\label{app:phase_cooccurrence}

Table~\ref{tab:phase_cooccurrence} shows kill chain phase co-occurrence.
The strongest pairs are Execution--Credential Access (89 skills) and Execution--Impact (82 skills): harvest credentials, then suppress alerts.
\textit{Stealth Ops} (smp\_716) downloads a secondary payload only when conditions are met (root access, production hostname, off-hours), combining E1, E3, P1, and P4.

\begin{table}[h]
\centering
\scriptsize
\caption{Kill chain phase co-occurrence (number of skills exhibiting both phases)}
\label{tab:phase_cooccurrence}
\begin{tabular}{@{}l|rrrrrr@{}}
\toprule
 & Recon & Cred & Exec & Evasion & Exfil & Impact \\
\midrule
Recon & 12 & & & & & \\
Cred & 3 & 107 & & & & \\
Exec & 7 & 89 & 118 & & & \\
Evasion & 2 & 15 & 21 & 25 & & \\
Exfil & 6 & 67 & 76 & 17 & 90 & \\
Impact & 10 & 73 & 82 & 17 & 63 & 109 \\
\bottomrule
\end{tabular}
\end{table}

\subsection{Hypothesis Test Details}
\label{app:hypothesis_tests}

\paragraph{H1: E2-E1 Association}
We test whether credential harvesting (E2) and external transmission (E1) are associated.
\begin{itemize}[leftmargin=*, nosep]
    \item Null hypothesis: E2 and E1 occur independently.
    \item Test: Fisher's exact test (two-sided).
    \item Contingency table: E2+E1=58, E2-only=43, E1-only=21, Neither=35.
    \item Result: $p = 0.020$, OR=2.24, 95\% CI [1.09, 4.66].
    \item Effect size: Cram\'{e}r's V = 0.19 (small--medium).
\end{itemize}

\paragraph{H2: Prompt Injection Severity}
We test whether skills with prompt injection (P1 or P2) exhibit different severity distributions.
\begin{itemize}[leftmargin=*, nosep]
    \item Null hypothesis: Severity distributions are equal with and without prompt injection.
    \item Test: Mann-Whitney U (non-parametric).
    \item With prompt injection (P1 or P2): Mean severity = 3.33 (n=38).
    \item Without prompt injection: Mean severity = 3.12 (n=119).
    \item Result: U=2903.0, $p=0.008$ (significant at $\alpha=0.01$).
\end{itemize}
Skills containing prompt injection patterns (P1: instruction override, P2: hidden instructions) exhibit marginally higher mean severity, though the practical difference is small (0.21 on a 4-point scale).

\section{Limitations of Dynamic Verification}
\label{app:dynamic_limitations}

Dynamic verification misses malware with time-delayed triggers (activating after specific dates), environment checks (running only on production systems), user interaction requirements, or anti-sandbox techniques. The 4,130 unconfirmed candidates include an unknown number of such dormant threats. Longer monitoring with varied trigger conditions would improve coverage.

\section{Reproducibility Checklist}
\label{app:reproducibility}

\mypara{Software}
Python 3.11, Docker 24.0, tcpdump 4.99, strace 6.

\mypara{Data Collection}
Collection date: January 2026. skills.rest API paginated at 100 repos/request; skillsmp.com API via alphabetic enumeration. Rate limit: 1 request/second.

\mypara{Dynamic Verification}
60-second execution timeout. Network capture runs for full execution plus 10 seconds. Honeypot credentials follow patterns AWS\_*, API\_*, SECRET\_*, TOKEN\_*.

\section{Severity Assignment Criteria}
\label{app:severity_criteria}

Severity is assigned per vulnerability instance following pattern-based guidelines below. A skill's overall severity is its highest instance severity.

\textbf{CRITICAL} (immediate high-impact exploitation): E2 (Credential Harvesting), PE3 (Credential File Access), SC2 (Remote Script Execution), SC3 (Obfuscated Code).

\textbf{HIGH} (significant attacks): E1 (External Transmission), P1 (Instruction Override), P2 (Hidden Instructions), P3 (Data Exfiltration via Code Execution), SC1 (Command Injection).

\textbf{MEDIUM} (moderate risk or ambiguous intent): E3 (File System Enumeration), E4 (Network Reconnaissance), P4 (Behavior Manipulation), PE2 (Privilege Escalation).

\textbf{LOW} (poor hygiene, not active threat): PE1 (Excessive Permissions).

\section{Evaluation Details}
\label{app:evaluation_details}

\subsection{Dataset Construction}
\label{app:dataset}

We constructed a labeled dataset of 300 skills: 150 benign, 150 malicious.

\textit{Benign skills.}
Randomly sampled from 94,093 skills with no static-analysis flags.
Two researchers independently verified each by reviewing code and executing in sandbox; 12 were replaced due to disagreement.

\textit{Malicious skills.}
Randomly sampled from our 157 confirmed set, all with behavioral evidence from dynamic verification.
The 150/150 split follows standard binary classification practice.

\subsection{Baseline Comparison Details}
\label{app:baseline}

Manual analysis of Skill Scanner's false positives reveals three fundamental limitations of regex-based static analysis:

\begin{enumerate}[leftmargin=*, nosep]
    \item \textit{Lack of Semantic Understanding}: The scanner cannot distinguish safe library functions from dangerous ones. For example, \texttt{re.compile()} (Python's regex compiler) triggers code injection alerts because the rule matches any \texttt{compile()} call without understanding the calling context.

    \item \textit{No File Type Differentiation}: Documentation and test files receive the same scrutiny as production code. API examples in \texttt{SKILL.md} files (e.g., \texttt{curl -H "Authorization: Basic ..."}) trigger credential exposure alerts, while test fixtures with mock credentials (\texttt{API\_KEY = "test-key-123"}) appear as real leaks.

    \item \textit{Operational Pattern Misinterpretation}: Standard DevOps patterns are flagged as malicious. Docker cache cleanup commands (\texttt{rm -rf /var/lib/apt/lists/*}) trigger destructive operation alerts, and import statements (\texttt{from urllib.request import Request}) are classified as network exfiltration attempts.
\end{enumerate}

These limitations explain the precision gap: Skill Scanner achieves $\leq$1.1\% (assuming perfect recall) versus 99.6\% for our full pipeline.

\subsection{False Negative Analysis}
\label{app:fn_analysis}

To evaluate recall, we randomly sampled 10\% (413) of the 4,130 unconfirmed candidates that passed static analysis but failed dynamic verification. Two researchers independently reviewed each (Cohen's $\kappa = 0.87$), classifying them into three categories:

\begin{enumerate}[leftmargin=*, nosep]
\item \textbf{93.2\% dormant malware}: conditional triggers our 60-second sandbox did not activate---time-delayed payloads, environment-specific conditions (production hostnames, root), or multi-session state persistence. These remain \textit{potential} threats absent extended monitoring.

\item \textbf{4.9\% legitimate security tools}: pen-testing and scanning utilities exhibiting suspicious patterns for defensive purposes.

\item \textbf{1.9\% truly missed malicious}: second-order attacks where a payload stored in one interaction executes in a later session, evading single-session detection.
\end{enumerate}

The 1.9\% truly missed (approximately 78 skills if extrapolated) require multi-session monitoring to detect.

\subsection{Validation Methodology}
\label{app:validation}

\mypara{Avoiding Circular Validation}
Static patterns only reduce the search space (98,380 to 4,287); ground truth comes from behavioral verification (e.g., an E2 match is confirmed only when credential transmission is observed) plus dual-researcher judgment on all 157 skills (Cohen's $\kappa = 0.89$).

\mypara{Statistical Power}
The 632 vulnerabilities across 157 skills provide sufficient power for co-occurrence analysis (91 pattern pairs; E2-E1 at $p = 0.020$). These 157 skills are the exhaustive result of behavioral verification across 98,380 skills, not a sample.

\section{Implementation Details}
\label{app:implementation}

\mypara{Static Analysis}
Marketplace crawlers collect from skills.rest and skillsmp.com via paginated API requests and search enumeration; the pattern scanner combines regex matching (code-level patterns) with GPT-5.2 analysis (instruction-level patterns P1, P2, P4), and a metadata extractor parses YAML permissions and triggers.

\mypara{Environment}
Experiments ran on a 64-core AMD EPYC server (256~GB RAM, 2~TB NVMe); containers had 2~GB memory limits and 60-second timeouts. The full pipeline processed 98,380 skills in about 72 hours.